\documentclass{article}
\pdfoutput=1

\usepackage{arxiv}
\usepackage[utf8]{inputenc} 
\usepackage[T1]{fontenc} 
\usepackage{hyperref} 
\hypersetup{
colorlinks = true,
linkcolor = black,
anchorcolor = black,
citecolor = black,
filecolor = black,
urlcolor = blue
}
\usepackage{booktabs} 
\usepackage{amsfonts} 
\usepackage{nicefrac} 
\usepackage{microtype} 
\usepackage{amsmath,bm}
\usepackage{graphicx}
\usepackage[flushleft]{threeparttable}
\usepackage{multirow}
\usepackage{natbib}
\usepackage{makecell}
\usepackage{caption}
\usepackage{subcaption} 
\usepackage{float}

\newcommand{\figurehere}[1]{\begin{center}%
=========================\\%
Insert Figure #1 about here\\%
=========================\\%
\end{center}}
\newcommand{\tablehere}[1]{\begin{center}%
=========================\\%
Insert Table #1 about here\\%
=========================\\%
\end{center}}

\usepackage{array}
\newcommand{\PreserveBackslash}[1]{\let\temp=\\#1\let\\=\temp}
\newcolumntype{C}[1]{>{\PreserveBackslash\centering}p{#1}}
\newcolumntype{R}[1]{>{\PreserveBackslash\raggedleft}p{#1}}
\newcolumntype{L}[1]{>{\PreserveBackslash\raggedright}p{#1}}

\title{Extending Growth Mixture Model to Assess Heterogeneity in Joint Development with Piecewise Linear Trajectories in the Framework of Individual Measurement Occasions}

\author{
Jin Liu \thanks{CONTACT Jin Liu Email: Veronica.Liu0206@gmail.com, \textcircled{c}2022, American Psychological Association. This paper is not the copy of record and may not exactly replicate the final, authoritative version of the article. Please do not copy or cite without authors' permission. The final article will be available, upon publication, via its DOI: 10.1037/met0000500}\\
Biometrics Department\\
Vertex Pharmaceuticals\\
 \And
Robert A. Perera\\
Department of Biostatistics\\
Virginia Commonwealth University \\
}

\begin{document}
\maketitle
\begin{abstract}
Researchers continue to be interested in exploring the effects that covariates have on the heterogeneity in trajectories. The inclusion of covariates associated with latent classes allows for a more clear understanding of individual differences and a more meaningful interpretation of latent class membership. Many theoretical and empirical studies have focused on investigating heterogeneity in change patterns of a univariate repeated outcome and examining the effects on baseline covariates that inform the cluster formation. However, developmental processes rarely unfold in isolation; therefore, empirical researchers often desire to examine two or more outcomes over time, hoping to understand their joint development where these outcomes and their change patterns are correlated. This study examines the heterogeneity in parallel nonlinear trajectories and identifies baseline characteristics as predictors of latent classes. Our simulation studies show that the proposed model can tell the clusters of parallel trajectories apart and provide unbiased and accurate point estimates with target coverage probabilities for the parameters of interest in general. We illustrate how to apply the model to investigate the heterogeneity in the joint development of reading and mathematics ability from Grade K to $5$. In this real-world example, we also demonstrate how to select covariates that contribute the most to the latent classes and transform candidate covariates from a large set into a more manageable set with retaining the meaningful properties of the original set in the structural equation modeling framework.
\end{abstract}

\keywords{Growth Mixture Model \and Joint Development of Nonlinear Trajectories \and Feature Selection \and Feature Extraction \and Individual Measurement Occasions}

\section{Introduction}\label{intro}
\subsection{Motivating Example}
Multiple existing studies have examined the longitudinal records of mathematics achievement scores from the Early Childhood Longitudinal Study, Kindergarten Class (ECLS-K), with the hope of understanding the potential heterogeneity in trajectories and the possible causes. For example, \citet{Kohli2015PLGC1} and \citet{Liu2019BLSGMM} explored a random sample from ECLS-K and ECLS-K: 2011, respectively, and have demonstrated that the development of mathematics ability can be modeled as latent classes of trajectories. Additionally, \citet{Liu2019BLSGMM} examined how the baseline covariates, such as socioeconomic status and teacher-reported abilities, inform the latent class formation of nonlinear trajectories of mathematics development.

However, developmental processes rarely unfold in isolation. For instance, \citet{Peralta2020PBLSGM} and \citet{Liu2021PBLSGM} investigated a random subset from ECLS-K and ECLS-K: 2011, respectively, and have shown that the change patterns of the development of reading ability and mathematics ability are correlated over time. With these findings, it is of great interest to further explore the joint development of reading and mathematics ability to answer (1) whether there are latent classes of joint development of the two abilities, (2) if so, whether the latent classes of joint development are different from those of univariate developmental process, and (3) how the baseline characteristics, including demographic information, socioeconomic status, and teacher rating scales, inform the clusters of the joint development of the two abilities. To answer these questions, we propose a growth mixture model of joint development that includes covariates as predictors of class membership. Following these existing studies, we assume that both the mathematics and reading developmental trajectories take the bilinear spline functional form with an unknown knot due to theoretical and empirical considerations, which we will elaborate on later in the introduction section. 

\subsection{Growth Mixture Models for Joint Development}
Theoretical and empirical researchers widely utilize finite mixture models \citep{Muthen1999GMM} to explain heterogeneity in a sample through multiple and a finite number of probability distributions together in the sense of a linear combination. In practice, researchers usually assume that the within-class probability density function follows a normal distribution with a class-specific mean and variance, although these distributions may come from different or multiple different families in some circumstances. A growth mixture model (GMM) is a type of multivariate normal finite mixture model. The outcome matrix (i.e., the growth factors of individual trajectories) in the GMM is a mix of two or more latent subpopulations, each composed of their own multivariate normal distributions.

The GMM framework has received considerable attention over the past twenty years, with many studies examining its benefits. First, in a GMM, the within-class trajectories can take on almost any functional form of underlying change patterns, including parametric functions, such as linear, quadratic, and Jenss-Bayley growth curves, as well as nonparametric functions, such as spline growth curve. More importantly, the functional forms of trajectories are not necessarily the same across latent classes. For example, in a two-class GMM, trajectories in one class may take a quadratic functional form, whereas the curves in the other class may take a linear function where the mean and variance of the latent quadratic slope and quadratic-related covariances are zero. Moreover, the GMM is a probability-based approach with consideration of uncertainty, where posterior probabilities of class membership for each individual can be calculated. Additionally, researchers can utilize several statistical fit indices to decide the optimal model since the GMM is a model-based clustering method \citep{Nylund2007number}.

Although most studies of GMMs focused on a univariate repeated outcome, some theoretical studies have proposed multiple ways of approaching multiple longitudinal outcomes. For example, \citet{Zucker1995PGMM} constructed a joint mixed-effects model to deal with longitudinal data involving multiple response variables, with each outcome variable postulated to take a linear functional form. Alternatively, \citet{Putter2008PGMM} proposed a two-stage model to estimate the parameters. The estimates from the first stage include the class-specific mean vector and variance-covariance structure, plus the mixing proportions based on the first longitudinal outcome. The relation between the latent classes and the other outcome(s) and covariates are examined in the second step. In addition, \citet{Saebom2017PGMM} proposed a latent variable model that allows for examining underlying joint patterns of multiple latent class variables. Empirical researchers also desire to investigate the heterogeneity in joint development. For instance, \citet{Hix2004PGMM} employed a multivariate associative growth mixture model to analyze the heterogeneity in adolescent alcohol and marijuana use over time. In this article, following \citet{Zucker1995PGMM}, we apply the GMM framework to analyze joint development of multivariate repeated outcomes; that is, the submodel in each cluster is a multivariate growth model (MGM) \cite[Chapter~8]{Grimm2016growth}, also referred to as a parallel process and correlated growth model \citep{McArdle1988Multi}. Note that the major differences between the current study and \citet{Zucker1995PGMM} lie in that (1) the GMM for joint development in this study is constructed in the structural equation modeling (SEM) framework, and (2) each of longitudinal processes under investigation in this study postulated to take a nonlinear function, specifically, the bilinear spline functional form. 

\subsection{Introduction of Bilinear Spline Functional Form}
One of the essential aspects of modeling change patterns is to capture the trajectory shape accurately. In longitudinal processes, the change patterns exhibit a nonlinear relationship to time $t$ when there are periods where change is more rapid than in others. There are multiple functional forms, such as polynomial, Jenss-Bayley, or piecewise, to describe the nonlinear change patterns. The choice of the functional form is important when depicting nonlinear trajectories. Driven by theoretical considerations, researchers usually decide the functional form to estimate parameters with an interpretation that aligns with research questions directly \citep{Cudeck2007Nonlinear}. The choice can also be an empirical selection: fit a pool of candidate models with different functional forms and select the function that best describes the change patterns. 

Linear spline growth curve models \cite[Chapter~11]{Grimm2016growth}, also referred to as piecewise linear models \citep{Harring2006nonlinear, Kohli2011PLGC, Kohli2013PLGC1, Kohli2013PLGC2, Sterba2014individually, Kohli2015PLGC1}, are a statistical tool that allows for different growth rates corresponding to different phases of a developmental process. In this type of model, the change-points or `knots' at which the two segments join together must be determined. The knots can be specified by domain knowledge, for example, \citet{Dumenci2019knee, Flora2008knot}, or be estimated as unknown parameters as multiple existing studies such as \citet{Cudeck2003knot_F, Harring2006nonlinear, Kohli2011PLGC, Kohli2013PLGC1}. 

Theoretically, longitudinal processes usually consist of multiple stages in many domains.  For example, in developmental studies, lots of psychological and educational phenomena are comprised of different phases. In the biomedical area, the recovery process from surgery, such as knee arthroplasty, is also considered in two stages. One possible index to evaluate the recovery process is the pain score: patients first recover from the surgical pain and then get better gradually \citep{Dumenci2019knee}. This linear spline functional form is of particular interest in estimating stage-specific change rates of these longitudinal processes. Additionally, the interpretation of the estimated knot, which allows for understanding when the transition from one stage to the other occurs, is unique to a piecewise function. More importantly, with an assumption that each repeated outcome follows a segmented linear growth curve when exploring joint development, we obtain stage-specific associations between multiple trajectories over time \citep{Liu2021PBLSGM, Peralta2020PBLSGM}.

Empirically, multiple existing studies have demonstrated that the growth rate in mathematics and reading skills slow down in developmental processes. The linear spline functional form can capture the underlying developmental patterns of these two abilities and outperforms other nonlinear functional forms, such as polynomial and Jenss-Bayley growth curve \citep{Liu2021PBLSGM, Peralta2020PBLSGM, Kohli2015PLGC2, Kohli2017PLGC}. These studies have also shown that the outcome-specific knot is individually different in the joint development of reading and mathematics abilities, assuming that all individuals are from one population. If we relax the one population assumption, another possible reason for the heterogeneity in knots is that these change points are different across latent classes. 

Accordingly, we propose to utilize a parallel bilinear spline growth curve model (PBLSGM) with unknown fixed knots \citep{Liu2021PBLSGM} as the within-class model, assuming that the outcome-specific knot is roughly the same across all individuals in each class\footnote{It is possible to estimate both fixed and random effects of these class-specific knots. In this project, we only focus on the fixed effects since we want to build a relatively parsimonious model given that the mixture model for a joint development itself is complicated.}, to examine the motivating data. Similar to \citet{Liu2021PBLSGM}, we build the model in the framework of individual-measurement occasions with `definition variables' \citep{Mehta2000people, Mehta2005people}, which adjust model parameters to individual-specific values to avoid potential inadmissible estimation \citep{Blozis2008coding, Coulombe2015ignoring}. 

In addition to applying the GMM with the proposed within-class model to explore possible clusters of the joint development, we can also employ the GMM in other fields, where latent classes are theoretically defensible. For example, \citet{Robertsson2000knee, Baker2007knee} have shown that although the majority of patients benefited from knee arthroplasty, a proportion of patients reported persistent pain or functional deficiencies in a year following the surgery, suggesting that there are responders and non-responders to the surgery. Physical function scales and pain scales are two possible indices to evaluate the recovery process \citep{Dumenci2019knee}. The proposed GMM allows for investigating the heterogeneity in short-term and long-term joint recovery processes. 

\subsection{Implementation Challenges of Mixture Models for Joint Development}
Similar to the GMM with a univariate repeated outcome, implementing the GMM with multivariate repeated outcomes poses statistical challenges such as determining the optimal number of latent classes and deciding which covariates may inform class membership and how to add these covariates. For the GMM with a univariate repeated outcome, researchers usually conduct the enumeration process, which excludes any covariates, in an empirical approach: fitting a pool of candidate GMMs with different numbers of latent classes and selecting the `best' model and the optimal number of clusters via the Bayesian information criterion (BIC) \citep{Nylund2007number}. For the GMM with parallel repeated outcomes in the current study, we do not intend to develop a novel metric for choosing the number of latent classes. Instead, we can still follow the SEM literature convention and determine the optimal number of latent classes using the BIC from the statistical perspective. Alternatively, we can sometimes make this decision by answering a specific research question for the GMM with multivariate repeated outcomes in the scenario where the cluster information of univariate development is available.  

We also need to decide to include which covariates in mixture models, which is also challenging. First of all,  the candidate pool of independent variables in the educational and psychological domains where the GMM is widely employed is potentially huge. Some covariates are often highly correlated. In the statistics and machine learning literature, there are two common ways to shrink covariate space and address the potential collinearity issue: feature extraction and feature selection, which can be realized by the exploratory factor analysis (EFA) \citep{Spearman1904factor} and the structural equation model forests (SEM Forests) \citep{Brandmaier2016semForest} in the SEM framework. In this article, we follow \citet{Liu2019BLSGMM} and \citet{Liu2020MoE} to demonstrate how to utilize these two methods to deal with the candidate covariate set for the motivating example with an assumption that all independent variables only have indirect effects on the heterogeneity of joint development. 

In the GMMs, the inclusion of covariates to inform the class formation can be realized in a one-step approach \citep{Clogg1981one, Goodman1974one, Haberman1979one, Hagenaars1993one, Vermunt1997one, Bandeen1997one, Dayton1988one, Kamakura1994one, Yamaguchi2000one} or stepwise methods, including a two-step \citep{Bakk2017two, Liu2019BLSGMM} or a three-step approach \citep{Bolck2004three, Vermunt2010three, Asparouhov2014three}. Multiple existing studies have shown that the one-step model outperforms the stepwise methods in terms of bias, mean squared error (MSE), and coverage probability (CP). Accordingly, we decided to follow the recommendation in multiple recent studies and construct a one-step mixture model in a stepwise fashion \citep{Liu2020MoE}, also referred to as the adjusted one-step approach \citep{Kim2016expert, Hsiao2020mediation}. These recent studies recommend (1) conducting the enumeration process without any covariates to have a stable number of clusters and (2) constructing a model with covariates and the determined number to estimate all parameters. 

In the remainder of this article, we first describe the model specification and model estimation of the GMM with PBLSGM as the within-class model. We then depict the design of the Monte Carlo simulation for model evaluation. We evaluate how the proposed model works through the performance measures, including the relative bias, the empirical standard error (SE), the relative root-mean-squared-error (RMSE), and the empirical coverage for a nominal $95\%$ confidence interval of each parameter of interest. We also compare the accuracy of the proposed GMM to that of the GMM with a univariate repeated outcome. In the Application section, we analyze the motivating data, longitudinal reading and mathematics achievement scores from the Early Childhood Longitudinal Study, Kindergarten Class $2010-11$ (ECLS-K:2011)\footnote{We want to examine ECLS-K:2011 instead of ECLS-K because the former contains baseline teacher-rated students' behavior questions, including attentional focus and inhibitory control, whose effects are of our interest.}, to demonstrate how to shrink covariate space and construct the proposed GMMs. Finally, we present a broad discussion concerning practical considerations, methodological considerations as well as future directions.

\section{Method}\label{method}
\subsection{Bilinear Spline Growth Curve Model with a Fixed Knot}
In this section, we briefly describe the latent growth curve (LGC) model with a linear-linear piecewise functional form with a fixed knot, which is utilized to analyze a univariate change pattern and estimate a change-point with the assumption that the knot is roughly the same across all individuals. Each of the two stages takes a linear functional form in this model, and the two linear segments join at a change-point or a `knot'.  In the framework of individual measurement occasions, the measurement $y_{ij}$ at the $j^{th}$ time point of $i^{th}$ individual $t_{ij}$ is 
\begin{equation}\label{eq:fun}
y_{ij}=\begin{cases}
\eta^{[y]}_{0i}+\eta^{[y]}_{1i}t_{ij}+\epsilon^{[y]}_{ij} & t_{ij}\le\gamma^{[y]}\\
\eta^{[y]}_{0i}+\eta^{[y]}_{1i}\gamma^{[y]}+\eta_{2i}(t_{ij}-\gamma^{[y]})+\epsilon^{[y]}_{ij} & t_{ij}>\gamma^{[y]}\\
\end{cases}.
\end{equation}
\citet{Harring2006nonlinear} showed there are five parameters, including one intercept and slope for each linear piece, and a knot, in the linear-linear piecewise model, but the degrees-of-freedom of the bilinear spline is four as two linear pieces join at the knot. As shown in Equation (\ref{eq:fun}), we consider the initial status ($\eta^{[y]}_{0i}$), two slopes ($\eta^{[y]}_{1i}$ and $\eta^{[y]}_{2i}$), and the knot ($\gamma^{[y]}$) as the four free parameters in the current study, where $\eta^{[y]}_{0i}$, $\eta^{[y]}_{1i}$ and $\eta^{[y]}_{2i}$ are individual-level while $\gamma^{[y]}$ is population-level (or subpopulation-level when modeling latent classes of trajectories). All four parameters determine the change-pattern of the growth curve of $\boldsymbol{y}_{i}$.

The piecewise function defined in Equation (\ref{eq:fun}) cannot be specified directly in an existing SEM software such as \textit{Mplus} and the \textit{R} package \textit{OpenMx} because a conditional statement is not allowed when specifying a model. Accordingly, we have to reparameterize growth factors to unify pre- and post-knot expressions, which can be realized in multiple ways, as presented in earlier studies. For example, \citet{Harring2006nonlinear} proposed to reparameterize the initial status and two slopes to the average of the two intercepts, the average of the two slopes, and the half difference between the two slopes. Alternatively, \citet[Chapter~11]{Grimm2016growth} suggested reexpressing the three original growth factors (i.e., $\eta^{[y]}_{0i}$, $\eta^{[y]}_{1i}$ and $\eta^{[y]}_{2i}$) as the measurement of the knot and two slopes. In addition, \citet{Liu2019BLSGM} reparameterized the three growth factors as the measurement of the knot, the average of the two slopes, and the half difference between the two slopes. 

Note that the only benefit of reparameterization lies in that it allows for specifying the model in Equation (\ref{eq:fun}). Although the model with original growth factors and reparameterized growth factors are mathematically equivalent, the reparameterized coefficients may no longer be directly related to the underlying developmental process and therefore lack meaningful and substantive interpretation. From this perspective, the approach proposed in \citet[Chapter~11]{Grimm2016growth} looks promising since all reparameterized coefficients are still directly related to the growth patterns. However, we need to call the \textit{minimum} and \textit{maximum} functions to specify the model in Equation (\ref{eq:fun}) with this method. Only the \textit{R} package \textit{OpenMx} allows for these two functions currently, which means it is impossible to apply this approach in \textit{Mplus}. Accordingly, in this study, we follow the reparameterized method in \citet{Liu2019BLSGM} because the transformation between the original and the reparameterized growth factors for joint development is well documented in \citet{Liu2021PBLSGM} and ready to use. We then write the repeated outcome as
\begin{equation}\label{eq:uni}
\boldsymbol{y}_{i}=\boldsymbol{\Lambda}_{i}^{[y]}\times\boldsymbol{\eta}^{[y]}_{i}+\boldsymbol{\epsilon}^{[y]}_{i},
\end{equation}
where 
\begin{equation}\nonumber
\boldsymbol{\eta}_{i}^{[y]} = \left(\begin{array}{rrr}
\eta^{'[y]}_{0i} & \eta^{'[y]}_{1i} & \eta^{'[y]}_{2i} 
\end{array}\right)^{T}
= \left(\begin{array}{rrr}
\eta^{[y]}_{0i}+\gamma^{[y]}\eta^{[y]}_{1i} & \frac{\eta^{[y]}_{1i}+\eta^{[y]}_{2i}}{2} & \frac{\eta^{[y]}_{2i}-\eta^{[y]}_{1i}}{2} 
\end{array}\right)^{T}
\end{equation}
and
\begin{equation}\nonumber
\begin{aligned}
&\boldsymbol{\Lambda}_{i}^{[y]} = \left(\begin{array}{rrr}
1 & t_{ij}-\gamma^{[y]} & |t_{ij}-\gamma^{[y]}| 
\end{array}\right)
&(j=1,\cdots, J).
\end{aligned}
\end{equation}
We provide the detailed deviation of the reparameterized growth factors and corresponding factor loadings in the Online Supplementary Document. The LGC with bilinear spline functional form can be extended to the GMM framework \citep{Kohli2011PLGC, Kohli2013PLGC1, Liu2019BLSGMM} or MGM framework \citep{Liu2021PBLSGM}. In the following sections, we demonstrate how to extend this model to the scenario with multivariate repeated outcomes and subpopulations (i.e., latent classes). 

\subsection{Model Specification of Growth Mixture Model with Parallel Bilinear Spline Growth Curves with Fixed Knots}\label{method:spec}
In this section, we specify the GMM with a parallel bilinear growth curve model (PBLSGM) with unknown fixed knots as the within-class model to investigate the heterogeneity of joint development and its possible causes. Suppose we have bivariate growth curves of repeated outcomes $\boldsymbol{y}_{i}$ and $\boldsymbol{z}_{i}$ for each individual and $K$ pre-specified number of latent classes, for $k=1$ to $K$ latent classes and $i=1$ to $n$ individuals, we express the model as 
\begin{align}
&p(\boldsymbol{y}_{i},\boldsymbol{z}_{i} |c_{i}=k,\boldsymbol{x}_{i})=\sum_{k=1}^{K}\pi(c_{i}=k|\boldsymbol{x}_{i})\times p(\boldsymbol{y}_{i},\boldsymbol{z}_{i}|c_{i}=k),\label{eq:GMM}\\
&\pi(c_{i}=k|\boldsymbol{x}_{i})=\begin{cases}
\frac{1}{1+\sum_{k=2}^{K}\exp(\beta_{0}^{(k)}+\boldsymbol{\beta}^{(k)T}\boldsymbol{x}_{i})} & \text{Reference Group ($k=1$)}\\
\frac{\exp(\beta_{0}^{(k)}+\boldsymbol{\beta}^{(k)T}\boldsymbol{x}_{i})} {1+\sum_{k=2}^{K}\exp(\beta_{0}^{(k)}+\boldsymbol{\beta}^{(k)T}\boldsymbol{x}_{i})} & \text{Other Groups ($k=2,\dots, K$)}
\end{cases},\label{eq:gating}\\
&\begin{pmatrix}
\boldsymbol{y}_{i} \\ \boldsymbol{z}_{i}
\end{pmatrix}|(c_{i}=k)=
\begin{pmatrix}
\boldsymbol{\Lambda}_{i}(\gamma^{[y]}) & \boldsymbol{0} \\ \boldsymbol{0} & \boldsymbol{\Lambda}_{i}(\gamma^{[z]})
\end{pmatrix}\times
\begin{pmatrix}
\boldsymbol{\eta}^{[y]}_{i} \\ \boldsymbol{\eta}^{[z]}_{i}
\end{pmatrix}|(c_{i}=k)+
\begin{pmatrix}
\boldsymbol{\epsilon}^{[y]}_{i} \\ \boldsymbol{\epsilon}^{[z]}_{i}
\end{pmatrix}|(c_{i}=k),\label{eq:expert1}\\
&\begin{pmatrix}
\boldsymbol{\eta}^{[y]}_{i} \\ \boldsymbol{\eta}^{[z]}_{i}
\end{pmatrix}|(c_{i}=k)=
\begin{pmatrix}
\boldsymbol{\mu_{\eta}}^{(k)[y]} \\ \boldsymbol{\mu_{\eta}}^{(k)[z]}
\end{pmatrix}+
\begin{pmatrix} 
\boldsymbol{\zeta}^{[y]}_{i} \\ \boldsymbol{\zeta}^{[z]}_{i}
\end{pmatrix}|(c_{i}=k).\label{eq:expert2}
\end{align}
Equation (\ref{eq:GMM}) defines a GMM that combines mixing proportions, $\pi(c_{i}=k|\boldsymbol{x}_{i})$, and within-class models, $p(\boldsymbol{y}_{i}, \boldsymbol{z}_{i}|c_{i}=k)$. In Equation (\ref{eq:GMM}), $\boldsymbol{x}_{i}$, $(\boldsymbol{y}_{i}, \boldsymbol{z}_{i})$ and $c_{i}$ are the covariates, bivariate repeated outcomes, and membership of the $i^{th}$ individual, respectively. We assume that $\boldsymbol{y}_{i}$ and $\boldsymbol{z}_{i}$ are $J\times1$ vectors, in which $J$ is the number of measurements. There are two constraints in Equation (\ref{eq:GMM}): $0\le \pi(c_{i}=k|\boldsymbol{x}_{i})\le 1$ and $\sum_{k=1}^{K}\pi(c_{i}=k|\boldsymbol{x}_{i})=1$. With Equation (\ref{eq:gating}), which defines mixing components as logistic functions of covariates $\boldsymbol{x}_{i}$, we allow for an association between the covariates and class membership. In Equation (\ref{eq:gating}), $\beta_{0}^{(k)}$ and $\boldsymbol{\beta}^{(k)}$ are logistic coefficients. 

Equations (\ref{eq:expert1}) and (\ref{eq:expert2}) together define the submodel in each latent class. Equation (\ref{eq:expert1}) expresses the bivariate repeated outcomes $(\boldsymbol{y}_{i},\boldsymbol{z}_{i})^{T}$ as a linear combination of growth factors. When the outcome-specific functional form is bilinear spline growth curve with an unknown fixed knot, Equation (\ref{eq:expert1}) can be viewed as an extension of Equation (\ref{eq:uni}) with joint development. Accordingly, $\boldsymbol{\eta}^{[u]}_{i}(u=y,z)$ is a $3\times1$ vector of outcome-specific growth factors and $\boldsymbol{\Lambda}_{i}(\gamma^{[u]})$ is a $J\times3$ matrix of corresponding factor loadings. Additionally, $\boldsymbol{\epsilon}^{[u]}_{i}$ is a $J\times 1$ vector of outcome-specific residuals of the $i^{th}$ individual. Equation (\ref{eq:expert2}) further expresses the growth factors as deviations from their class-specific means. In the equation, $\boldsymbol{\mu_{\eta}}^{(k)[u]}$ is a $3\times 1$ vector of the outcome-specific growth factor means in the $k^{th}$ latent class and $\boldsymbol{\zeta}^{[u]}_{i}$ is a $3\times 1$ vector of the outcome-specific residual deviations from the mean vector of the $i^{th}$ individual. With the assumption that the class-specific growth factors of bivariate repeated outcomes follow a multivariate Gaussian distribution, the vector $\begin{pmatrix} \boldsymbol{\zeta}^{[y]}_{i} & \boldsymbol{\zeta}^{[z]}_{i}\end{pmatrix}^{T}|(c_{i}=k)$ can be further expressed as
\begin{equation}\label{eq:expert3}
\begin{pmatrix} 
\boldsymbol{\zeta}^{[y]}_{i} \\ \boldsymbol{\zeta}^{[z]}_{i}
\end{pmatrix}|(c_{i}=k)\sim \text{MVN}\bigg(\boldsymbol{0}, 
\begin{pmatrix}
\boldsymbol{\Psi}_{\boldsymbol{\eta}}^{(k)[y]} & \boldsymbol{\Psi}_{\boldsymbol{\eta}}^{(k)[yz]} \\
& \boldsymbol{\Psi}_{\boldsymbol{\eta}}^{(k)[z]}
\end{pmatrix}\bigg),
\end{equation}
where $\boldsymbol{\Psi}_{\boldsymbol{\eta}}^{(k)[u]}$ is a $3\times 3$ variance-covariance matrix of the outcome-specific growth factors and $\boldsymbol{\Psi}_{\boldsymbol{\eta}}^{(k)[yz]}$ is a $3\times 3$ matrix of the between-construct growth factor covariances in the $k^{th}$ latent class. To simplify the model, we assume that the outcome-specific residuals ($\boldsymbol{\epsilon}^{[u]}_{i}$) in Equation (\ref{eq:expert1}) are independent and identically normally distributed over time, and the class-specific residual covariances are homogeneous over time, that is,
\begin{equation}\nonumber
\begin{pmatrix} 
\boldsymbol{\epsilon}^{[y]}_{i} \\ \boldsymbol{\epsilon}^{[z]}_{i}
\end{pmatrix}|(c_{i}=k)\sim \text{MVN}\bigg(\boldsymbol{0}, 
\begin{pmatrix}
\theta^{(k)[y]}_{\epsilon}\boldsymbol{I} & \theta^{(k)[yz]}_{\epsilon}\boldsymbol{I} \\
& \theta^{(k)[z]}_{\epsilon}\boldsymbol{I}
\end{pmatrix}\bigg),
\end{equation}
where $\boldsymbol{I}$ is a $J\times J$ identity matrix. As stated earlier, the reparameterized growth factors are no longer related to the underlying change patterns and need to be transformed back for interpretation purposes. We also extend the (inverse-)transformation functions and matrices for the reduced model in \citet{Liu2021PBLSGM} to the GMM framework. Through inverse-transformation, we can obtain the coefficients directly related to the underlying developmental processes, and therefore, meaningful and substantive interpretation. We provide the details of the class-specific (inverse-) transformation in the Online Supplementary Document. 

\subsection{Model Estimation}\label{method:est}
In this section, we demonstrate how to estimate the parameters of interest from the proposed mixture model. We first write the within-class model implied mean vector and variance-covariance matrix of the bivariate repeated outcomes for the $i^{th}$ individual in the $k^{th}$ unobserved group as
\begin{equation}\label{eq:mean}
\boldsymbol{\mu}_{i}^{(k)}=\begin{pmatrix}
\boldsymbol{\mu}^{(k)[y]}_{i} \\ \boldsymbol{\mu}^{(k)[z]}_{i}
\end{pmatrix}=\begin{pmatrix}
\boldsymbol{\Lambda}_{i}(\gamma^{[y]}) & \boldsymbol{0} \\ \boldsymbol{0} & \boldsymbol{\Lambda}_{i}(\gamma^{[z]})
\end{pmatrix}\times\begin{pmatrix}
\boldsymbol{\mu}^{(k)[y]}_{\boldsymbol{\eta}} \\ \boldsymbol{\mu}^{(k)[z]}_{\boldsymbol{\eta}}
\end{pmatrix}
\end{equation}
and
\begin{equation}\label{eq:var}
\begin{aligned}
\boldsymbol{\Sigma}^{(k)}_{i}&=\begin{pmatrix}
\boldsymbol{\Sigma}^{(k)[y]}_{i} & \boldsymbol{\Sigma}^{(k)[yz]}_{i} \\
& \boldsymbol{\Sigma}^{(k)[z]}_{i}
\end{pmatrix}\\
&=\begin{pmatrix}
\boldsymbol{\Lambda}_{i}(\gamma^{[y]}) & \boldsymbol{0} \\ \boldsymbol{0} & \boldsymbol{\Lambda}_{i}(\gamma^{[z]})
\end{pmatrix}\times\begin{pmatrix}
\boldsymbol{\Psi}^{(k)[y]}_{\boldsymbol{\eta}} & \boldsymbol{\Psi}^{(k)[yz]}_{\boldsymbol{\eta}} \\
& \boldsymbol{\Psi}^{(k)[z]}_{\boldsymbol{\eta}}
\end{pmatrix}\times\begin{pmatrix}
\boldsymbol{\Lambda}_{i}(\gamma^{[y]}) & \boldsymbol{0} \\ \boldsymbol{0} & \boldsymbol{\Lambda}_{i}(\gamma^{[z]})
\end{pmatrix}^{T}+\begin{pmatrix}
\theta^{(k)[y]}_{\epsilon}\boldsymbol{I} & \theta^{(k)[yz]}_{\epsilon}\boldsymbol{I} \\
& \theta^{(k)[z]}_{\epsilon}\boldsymbol{I}
\end{pmatrix}.
\end{aligned}
\end{equation}

We then estimate the class-specific parameters and logistic coefficients for the GMM specified in Equations (\ref{eq:GMM}), (\ref{eq:gating}), (\ref{eq:expert1}) and (\ref{eq:expert2}). The parameters that need to be estimated include
\begin{equation}\label{eq:theta}
\begin{aligned}
\boldsymbol{\Theta}=&\{\boldsymbol{\mu}^{(k)[u]}_{\boldsymbol{\eta}}, \boldsymbol{\Psi}^{(k)[u]}_{\boldsymbol{\eta}}, \boldsymbol{\Psi}^{(k)[yz]}_{\boldsymbol{\eta}}, \theta^{(k)[u]}_{\epsilon}, \theta^{(k)[yz]}_{\epsilon}, \beta_{0}^{(k)}, \boldsymbol{\beta}^{(k)}\}\\
=&\{\mu^{(k)[u]}_{\eta_{0}}, \mu^{(k)[u]}_{\eta_{1}}, \mu^{(k)[u]}_{\eta_{2}}, \gamma^{(k)[u]}, \psi^{(k)[u]}_{00}, \psi^{(k)[u]}_{01}, \psi^{(k)[u]}_{02}, \psi^{(k)[u]}_{11}, \psi^{(k)[u]}_{12}, \psi^{(k)[u]}_{22},\\
&\psi^{(k)[yz]}_{00}, \psi^{(k)[yz]}_{01}, \psi^{(k)[yz]}_{02}, \psi^{(k)[yz]}_{10}, \psi^{(k)[yz]}_{11}, \psi^{(k)[yz]}_{12}, \psi^{(k)[yz]}_{20}, \psi^{(k)[yz]}_{21}, \psi^{(k)[yz]}_{22}, \ \ \ \ \ \ \ \ \ \ \ \\
& \theta^{(k)[u]}_{\epsilon}, \theta^{(k)[yz]}_{\epsilon}, \beta_{0}^{(k)}, \boldsymbol{\beta}^{(k)}\}\\
&u=y,\ z,\\
&k=2,\dots,K \text{ for } \beta_{0}^{(k)}, \boldsymbol{\beta}^{(k)},\\
&k=1,\dots,K \text{ for other parameters}.
\end{aligned}
\end{equation}

We estimate $\boldsymbol{\Theta}$ using the full information maximum likelihood (FIML) method to account for the possible heterogeneity in individual contributions to the likelihood function. The log-likelihood function of the model specified in Equations (\ref{eq:GMM}), (\ref{eq:gating}), (\ref{eq:expert1}) and (\ref{eq:expert2}) can be expressed as
\begin{equation}\nonumber
\begin{aligned}
\log lik(\boldsymbol{\Theta})&=\sum_{i=1}^{n}\log\bigg(\sum_{k=1}^{K}\pi(c_{i}=k|\boldsymbol{x}_{i})\times p(\boldsymbol{y}_{i},\boldsymbol{z}_{i}|c_{i}=k)\bigg)\\
&=\sum_{i=1}^{n}\log\bigg(\sum_{k=1}^{K}\pi(c_{i}=k|\boldsymbol{x}_{i})\times p(\boldsymbol{y}_{i},\boldsymbol{z}_{i}|\boldsymbol{\mu}_{i}^{(k)},\boldsymbol{\Sigma}_{i}^{(k)})\bigg).
\end{aligned}
\end{equation}
In the current study, the proposed GMM is built using the R package \textit{OpenMx} with CSOLNP optimizer \citep{Pritikin2015OpenMx, OpenMx2016package, User2020OpenMx, Hunter2018OpenMx}, which allows for matrix calculations so that we can implement the class-specific inverse-transformation function and matrix detailed in the Online Supplementary Document efficiently. In the online appendix (\url{https://github.com/Veronica0206/Extension_projects}), we provide the \textit{OpenMx} code for the proposed model and a demonstration. \textit{Mplus} 8 syntax is also provided for the proposed model in the online appendix for researchers who are interested in using \textit{Mplus}.

\section{Model Evaluation}\label{Evaluation}
In this section, we evaluate the proposed GMM with a PBLSGM as the within-class model using a Monte Carlo simulation study with two goals. The first goal is to assess how the proposed model performs when the two repeated outcomes are correlated over time. We evaluate the model through performance measures, including the relative bias, empirical standard error (SE), relative root-mean-square error (RMSE), and empirical coverage for a nominal $95\%$ confidence interval of each parameter. We list the definitions and estimates of the four performance metrics in Table \ref{tbl:metric}. We also want to examine how well the proposed mixture model can distinguish trajectory clusters. Since having true membership in the simulation study, we employ the metric accuracy, which is defined as the fraction of all correctly classified instances, to assess how the model separates samples into `correct' clusters \cite[Chapter~1]{Bishop2006pattern}.

\tablehere{1}

To calculate the accuracy, we first obtain the posterior probabilities that indicate each individual belongs to the $k^{th}$ latent class, which can be realized by Bayes' theorem
\begin{equation}\nonumber
p(c_{i}=k)=\frac{\pi(c_{i}=k|\boldsymbol{x}_{i})p(\boldsymbol{y}_{i}, \boldsymbol{z}_{i}|\boldsymbol{\mu}_{i}^{(k)},\boldsymbol{\Sigma}_{i}^{(k)})}{\sum_{k=1}^{K}\pi(c_{i}=k|\boldsymbol{x}_{i})p(\boldsymbol{y}_{i}, \boldsymbol{z}_{i}|\boldsymbol{\mu}_{i}^{(k)},\boldsymbol{\Sigma}_{i}^{(k)})}, 
\end{equation}
and assign each individual to the class with the highest posterior probability of which that individual most likely belongs. Following \citet{McLachlan2000FMM}, we broke the tie among competing classes randomly when their posterior probabilities were equal to the maximum value. The second goal is to compare the performance metrics and accuracy obtained from the clustering algorithm on the bivariate repeated outcomes to those from the GMM with univariate development, hoping to explore whether the algorithm performance would be improved when it works on bivariate repeated outcomes. 

Guided by \citet{Morris2019simulation}, we decided the number of replications $S=1,000$ in the simulation study using an empirical method. In a pilot simulation study, standard errors of all parameters except the intercept variances were less than $0.15$. To keep the Monte Carlo standard error of the bias\footnote{$\text{Monte Carlo SE(Bias)}=\sqrt{Var(\hat{\theta})/S}$ \citet{Morris2019simulation}.} (the most important performance metric) below $0.005$, we needed at least $900$ repetitions. We then proceeded $S=1,000$ for more conservative consideration.

\subsection{Design of Simulation Study}\label{Evaluation:design}
We list all conditions that we considered in the simulation design in Table \ref{tbl:simu}. We fixed several conditions, including the sample size, the number of latent classes, the variance-covariance matrix of the outcome-specific growth factors in each latent class, the number of repeated measurements, and the time-window of individual measurement occasions, which are not the primary interest in this study. For example, we selected ten scaled and equally spaced waves as \citet{Liu2021PBLSGM} has shown that the parallel bilinear growth curve models performed decently in terms of the four performance measures and the shorter study duration (i.e., six scaled and equally spaced waves) only affected the model performance slightly. Similar to \citet{Liu2021PBLSGM}, the time-window of individual measurement occasions was set to be a medium level ($-0.25, +0.25$) around each wave \citep{Coulombe2015ignoring}. The variance-covariance structure of the growth factors usually varies with the time scale and measurement scale; we fixed it and kept the index of dispersion ($\sigma^{2}/\mu$) of each growth factor at a one-tenth scale \citep{Bauer2003GMM, Kohli2011PLGC, Kohli2015PLGC1}. Additionally, the correlations between outcome-specific growth factors in each latent class were set to be a moderate level ($\rho^{(k)[u]}=0.3$).

\tablehere{2}

The characteristic of the greatest importance for a model-based clustering method is how well the model can detect heterogeneity in samples and estimate class-specific parameters. The major condition hypothesized to influence such performance is the separation between latent classes, which is gauged by the difference in knot locations and the Mahalanobis distance between class-specific growth factors \citep{Kohli2015PLGC1, Liu2019BLSGM, Liu2020MoE} in this project. We set $0.50$, $0.75$ and $1.00$ as a small, medium, and large difference in outcome-specific knot locations in the simulation design\footnote{In the simulation design of multiple existing studies, the three levels were set to be $1.00$, $1.50$ and $2.00$ for the GMM with a univariate repeated outcome \citep{Kohli2015PLGC1, Liu2019BLSGMM, Liu2020MoE}.}.

In this study, we kept the within-construct Mahalanobis distance as $0.86$ (a small distance as in \citet{Kohli2015PLGC1}). Since the proposed model is for joint development, how the correlation between two trajectories affects model performance is worth exploring. We considered three levels of the between-construct growth factor correlation, $\pm0.3$ and $0$, for two considerations. On the one hand, the value of the between-construct growth factor correlation can help adjust Mahalanobis distance of the bivariate repeated outcomes slightly. Specifically, the Mahalanobis distance is $1.22$, $1.18$ and $1.35$ when the correlation is $0$, $+0.3$ and $-0.3$, respectively. On the other hand, with $0$ and $\pm{0.3}$ correlation conditions, it was of interest to investigate how the zero or positive (negative) moderate correlation affects the model performance. 

The allocation ratio that is roughly controlled by the intercept coefficient ($\beta_{0}$) in the logistic function was set as $1$:$1$ ($\beta_{0}=0$) or $1$:$2$ ($\beta_{0}=0.775$). The class mixing proportion $1$:$1$ was selected as it is a balanced allocation, while the other condition $1$:$2$ was chosen as we were interested in examining whether a more challenging condition in mixing proportions would affect the performance measures and accuracy. Additionally, we investigate several common change patterns, as shown in Table \ref{tbl:simu} (Scenario $1$, $2$ and $3$). We changed the knot location and the intercept mean for one repeated outcome (i.e., the mean trajectory of two latent classes were parallel) in all three scenarios. However, for the other repeated outcome, in addition to the knot location, we adjusted the mean values of the intercept, the first slope, and the second slope for the Scenario $1$, $2$, and $3$, respectively. We also considered two levels of residual variance ($1$ or $2$) to assess how the measurement precision affects the proposed model. 

\subsection{Data Generation and Simulation Step}\label{evaluation:step}
For each condition listed in Table \ref{tbl:simu}, we carried out the following two-step data generation for the proposed model. We obtained the membership $c_{i}$ from covariates $\boldsymbol{x}_{i}$ for each individual in the first step. We then generated the bivariate repeated outcomes $\boldsymbol{y}_{i}$ and $\boldsymbol{z}_{i}$ for each latent class simultaneously. The general steps are:
\begin{enumerate}
\item Obtain membership $c_{i}$ for the $i^{th}$ individual:
\vspace{-1mm}
\begin{enumerate}
\item Generate individual-level covariates $\boldsymbol{x}_{i}$,
\item Calculate the probability vector for each individual based on the covariates and a set of specified coefficients with a logit link, and assign each individual to the group with the highest probability,
\end{enumerate}
\vspace{-1mm}
\item Generate growth factors of bivariate repeated outcomes simultaneously with the prespecified mean vector and variance-covariance matrix for each latent class (listed in Table \ref{tbl:simu}) using the \textit{R} package \textit{MASS} \citep{Venables2002Statistics},
\vspace{-1mm}
\item Generate the scaled and equally-spaced time structure with ten repeated measures and obtain individual measurement occasions by allowing the time-window set as $t_{ij}\sim U(t_{j}-0.25, t_{j}+0.25)$ around each wave,
\vspace{-1mm}
\item Calculate factor loadings of each outcome for each individual from corresponding knot location and individual measurement occasions,
\vspace{-1mm}
\item Obtain the values of the bivariate repeated outcomes from the class-specific growth factors, corresponding factor loadings, class-specific knots and residual variances,
\vspace{-1mm}
\item Implement the proposed model on the generated data set, estimate the parameters, and construct corresponding $95\%$ Wald CIs, along with accuracy,
\vspace{-1mm}
\item Repeat steps $1$ through $6$ until achieving $1,000$ convergent solutions.
\end{enumerate}

\section{Results}\label{results}
\subsection{Model Convergence}
In this section, we investigate the convergence rate of the proposed GMM with a PBLSGM as the within-class model\footnote{In this study, we define \textit{convergence} as the solution with \textit{OpenMx} status code $0$ that suggests a successful optimization until up to $10$ runs with different sets of initial values \citep{OpenMx2016package}.}. When the two repeated outcomes are correlated over time, the model converged satisfactorily. Specifically, the proposed model's convergence rate was at least $94\%$ across the conditions with $\pm0.3$ between-construct growth factor correlation. The convergence rate in the scenarios with the large difference in knot locations (i.e., the difference in outcome-specific knot locations is $1.00$) was $100\%$. The worst condition had a non-convergence rate of $68/1068$, indicating that there were $68$ non-convergent datasets and required $1,068$ replications to reach $1,000$ converged solutions. It occurred under the condition with the unbalanced allocation (i.e., the ratio is $1$:$2$), the small difference (i.e., $0.5$) in outcome-specific knot locations, and the positive between-construct correlation (i.e., $\rho=0.3$, note that as mentioned earlier, the Mahalanobis distance of joint development is the smallest under the conditions with the positive between-construct correlation if the within-construct Mahalanobis distance is the same). Additionally, we noticed that the proposed model did not converge well when it was over-specified. Under the conditions with the zero between-construct correlation, where the GMM with joint development was not supposed to be utilized, the convergence rate was $54\%$-$77\%$. Due to the high non-convergence rate of these overspecified models, we do not evaluate their performance measures in the main text. 

\subsection{Performance Measures}
In this section, we present the simulation results of performance measures, including relative bias, empirical SE, relative RMSE, and empirical coverage probability of a nominal $95\%$ confidence interval for each parameter across all non-zero between-construct correlation conditions. In general, the proposed model is capable of providing unbiased and accurate point estimates with target coverage probabilities. We named the latent class with earlier knots as Class $1$ (i.e., the left cluster) while that with later knots as Class $2$ (i.e., the right cluster) in this section. Given the size of parameters and simulation conditions, we first calculated each performance measure across $1,000$ replications for each parameter under each condition. We then summarized the values of each performance metric from all the conditions under examination as the corresponding median and range for each parameter. The summary of each performance measure is provided in the Online Supplementary Document. 

The proposed model provided unbiased point estimates with small empirical SEs. Specifically, the magnitude of the relative biases of growth factor means, growth factor variances, between-construct growth factor covariances, and logistic coefficients were under $0.018$, $0.036$, $0.068$, and $0.097$, respectively. The magnitude of empirical SE of all parameters except intercept coefficients (i.e., intercept means, variances, and covariance) and logistic coefficients was under $0.29$. The empirical SEs of $\mu_{\eta0}^{(k)[u]}$, $\psi_{00}^{(k)[u]}$ and $\psi_{00}^{(k)[yz]}$ were around $0.50$, $3.00$, and $2.30$ respectively.

Moreover, the proposed model was capable of estimating parameters accurately. The magnitude of the relative RMSE of the growth factor means, growth factor variances, and growth factor covariance was under $0.14$, $0.29$, and $0.66$, respectively. The relative RMSE magnitude of the logistic coefficients was around $0.40$. 
Generally, the proposed GMM performed well in terms of empirical coverage probabilities of growth factor means, variances, and covariances since the median values of their coverage probabilities were around $0.92$. We noted that knots' coverage probabilities could be unsatisfactory. We then plotted the coverage probabilities of the class-specific knots stratified by the difference in the outcome-specific knot locations in Figure \ref{fig:KnotCP}. We noticed that the coverage probabilities of all knots were around $0.95$ under the conditions with large separation, although the coverage probabilities of knots, especially the $\boldsymbol{Y}$'s knot in Class $1$ and the $\boldsymbol{Z}$'s knot in Class $2$, were conservative under other conditions.

\figurehere{1}

To summarize, we generally obtained unbiased and accurate point estimates with target coverage probabilities from the proposed models. Some factors, such as separation in latent classes, influenced performance metrics. Specifically, greater separation improved model performance. Additionally, for the conditions with the unbalanced allocation (where the ratio is $1$:$2$), the performance metrics of the parameters in Cluster $2$ were better than those in Cluster $1$. It is not surprising given the larger sample size in the second latent class. Other conditions, such as the between-construct correlation magnitude or sign, did not affect the performance measures meaningfully (We provide the detailed summary of relative bias and empirical SE under the conditions with zero between-construct growth factor correlation in the Online Supplementary Document). 

\subsection{Accuracy}\label{R:Acc}
In this section, we evaluate accuracy across all conditions listed in Table \ref{tbl:simu}. We first calculated the mean of accuracy values over $1,000$ replications for each condition. We then plotted these mean values stratified by allocation ratio, separation in the outcome-specific knot locations, trajectory shapes, and the sign and magnitude of between-construct growth factor correlation, as shown in Figure \ref{fig:acc}. Generally, the mean value of accuracy was the greatest under the conditions with the large separation in outcome-specific knot locations (i.e., $1.00$), followed by the conditions with the medium separation (i.e., $0.75$) and then the small separation (i.e., $0.50$). The mean values of accuracy achieved $80\%$ when the difference was $1.00$.

\figurehere{2}

Moreover, the accuracy value under the conditions with negative between-construct correlations was greater than the value under the other conditions in general. It is not unexpected since the Mahalanobis distance between class-specific growth factors was relatively large when the correlation is negative, as stated earlier ($1.22$, $1.18$ and $1.35$ for $0$, $+0.3$ and $-0.3$ between-construct correlation when the within-construct Mahalanobis distance was fixed as $0.86$). Additionally, unbalanced allocation produced relatively higher accuracy, but trajectory shapes only affected the accuracy slightly. 

\subsection{Comparison to Models with Univariate Repeated Outcome}\label{R:compare}
In this section, we compare the GMM with joint development and those with univariate development concerning point estimates and accuracy. We noticed that the GMM with joint development outperformed the models for univariate development since the relative biases and empirical SEs of the joint development model were smaller than those from the univariate development models. We provide the median and range of relative bias and empirical SE of each parameter obtained from three models across the conditions with the large outcome-specific knot locations (i.e., $1.00$) in the Online Supplementary Document. Additionally, Figure \ref{fig:acc_compare} presents the mean accuracy of three models stratified by between-construct correlations, from which we observed that the accuracy of the joint development model was much higher than that of the univariate development model. 

\figurehere{3}

\section{Application}\label{application}
We have two goals in the application section. First, we demonstrate how to utilize the proposed GMM with a PBLSGM as the within-class model to analyze a real-world data set. Second, we extend two methods, the EFA and the SEM Forests, that can shrink the covariate space of the GMM with univariate development to the GMM with joint development. We extracted $500$ students randomly from ECLS-K: 2011 with complete records of repeated reading item response theory (IRT) scaled scores, mathematics IRT scaled scores, demographic information (sex, race/ethnicity, and age in months at each wave), socioeconomic status (including baseline family income and the highest education level between parents), baseline teacher-reported social skills (including self-control ability, interpersonal skills, externalizing problem and internalizing problem), baseline teacher-reported approach-to-learning, baseline teacher-reported children behavior question (including attentional focus and inhibitory control), and school information (baseline school type and location)\footnote{The total sample size of ECLS-K: 2011 is $n=18174$. After removing records with missing values (i.e., rows with any of NaN/-9/-8/-7/-1), the number of individuals is $n=1838$.}.

ECLS-K: 2011 is a nationwide longitudinal study of US children enrolled in about $900$ kindergarten programs beginning from the $2010-2011$ school year. In ECLS-K: 2011, children's reading and mathematics IRT scores were evaluated in nine waves: fall and spring of kindergarten, first and second grade, respectively, as well as spring of $3^{rd}$, $4^{th}$ and $5^{th}$ grade, respectively. Only about $30\%$ students were evaluated in $2011$ fall semester and $2012$ fall semester \citep{Le2011ECLS}. There are two time structures in the dataset, children's age (in months) and their grade-in-school. In this analysis, we used children's age (in months) to obtain individual measurement occasions. In the subset, $49.8\%$ and $50.2\%$ of students were boys and girls. Additionally, the sample was diverse with representations from White ($49.0\%$), Black ($5.2\%$), Latinx ($31.2\%$), Asian ($8.4\%$), and others ($6.2\%$). We dichotomized the variable race to be White ($49.0\%$) and others ($51.0\%$). At the start of the study, $10.8\%$ and $89.2\%$ of students were from private and public schools, respectively. All other covariates were treated as continuous variables in this analysis.

\subsection{Enumeration Process}\label{app:number}
This section demonstrates how to decide the number of latent classes of the GMM with joint development. Following the SEM literature convention, we select the model with the optimal number of classes without adding any covariates. We first fit one-, two- and three-class models for each univariate development and joint development. All nine models converged. Table \ref{tbl:compare} provides the estimated likelihood, information criteria (AIC and BIC), proportions of each latent class of each model, from which we can see that the optimal number of latent classes for the univariate development models and the joint development model was $3$ and $2$ determined by the BIC, respectively. This is not surprising. The BIC is a criterion that penalizes model complexity (i.e., the number of parameters), and the penalty of adding one latent class in the joint development model (that contains $32$ parameters in submodel) is much larger than that in the univariate development models (that have $11$ parameters in each latent class).

\tablehere{3}

Alternatively, the enumeration process of the joint development model can be driven by empirical knowledge or research interest. The optimal univariate development models suggest three clusters of either reading development trajectories or mathematics development trajectories in this application. Accordingly, a follow-up question of greater interest is to explore the joint development trajectories in three latent classes. Figures \ref{fig:traj_uni} and \ref{fig:traj_bi} present the model implied curves on the smooth lines that obtained from raw trajectories of each latent class of each ability obtained from the univariate development models and the joint development model, respectively. The estimated mixing proportions of the joint development changed: the mixing proportion of Class $1$ ($32.60\%$) increased while that of Class $2$ ($44.20\%$) and Class $3$ ($23.20\%$) decreased. Additionally, the margin of estimated trajectories of mathematics IRT scores in the pre-knot stage between Class $1$ and Class $2$  was wider in the joint development model. 

\figurehere{4}

\figurehere{5}

\subsection{Shrinking Covariate Space}\label{app:cov}
\citet{Liu2019BLSGMM} and \citet{Liu2020MoE} proposed to employ the EFA and the SEM Forests to shrink covariate space of the GMM with univariate development by conducting feature extraction and feature selection, respectively. Specifically, \citet{Liu2019BLSGMM} proposed to employ the EFA to address covariate spaces with high-dimension and highly correlated covariate subsets, which are common in psychological and educational domains. \citet{Liu2020MoE} recommended only including covariates that have great effects on the sample heterogeneity, which is decided by the output named `variable importance' of SEM Forests, in a GMM with univariate development. In this section, we extend the two methods to the GMM with joint development. 

\subsection*{Feature Extraction: Exploratory Factor Analysis}
We employ the EFA to address the potential collinearity issue for socioeconomic variables and teacher-reported ability/problem variables and derive an informative but non-redundant covariate set with fewer variables. Following \citet{Liu2019BLSGMM}, the EFA was conducted using the \textit{R} function \textit{factanal} in the \textit{stats} package \citep{Core2020stat}. Several criteria, including the eigenvalues greater than $1$ (EVG1) component retention criterion, scree test \citep{Cattell1966EFA, Cattell1967EFA}, and parallel analysis \citep{Horn1965EFA, Humphreys1969EFA, Humphreys1975EFA}, all suggested that two factors can explain the variance of socioeconomic variables and teacher-reported skills/problems. Additionally, we employed an orthogonal rotation, varimax, assuming that the factor of the socioeconomic variables and that of teacher-rated scores were independent. As a result, the first factor differentiates between teacher-reported skills and teacher-rated problems; the second factor can be interpreted as general socioeconomic status (the detailed output of the factor loadings and explained variance from the EFA is provided in the Online Supplementary Document). We then used the factor scores obtained by Bartlett's weighted least-squares methods \citep{Bartlett1937score} as well as sex, race/ethnicity, school type, and school location in the proposed model. 

\subsection*{Feature Selection: Structural Equation Model Forests}
Guided by \citet{Liu2020MoE}, we built SEM Forests for the univariate development models and the joint development model using the \textit{R} package \textit{semtree}. In addition to the original data set and the pool of candidate covariates, the input of the SEM Forests algorithm also includes a one-group model as the template model (i.e., a LGC for univariate development or a MGM for joint development). One output of SEM Forests is the variable importance scores in terms of predicting the model-implied mean vector and variance-covariance structure \citep{Brandmaier2016semForest}. For the parameter setting in this study, we used the bootstrapping sample method, $128$ trees and $2$ subsampled covariates at each node following \citet{Liu2020MoE}. 

The top four predictors of all three GMMs were parents' highest education, family income, attentional focus, and learning approach, although these variables' (relative) importance scores varied across models. We provide figures of variable importance scores of the three models in the Online Supplementary Document and will discuss the difference in the (relative) importance scores in the Discussion section. We decided to keep these four covariates and demographic information, including sex and race/ethnicity, as covariates in the proposed model. 

\subsection{Proposed Models}\label{app:model}
Tables \ref{tbl:GMM_selection} and \ref{tbl:GMM_reduction} present the estimates from the proposed GMM for joint development with covariates from feature selection and those from feature extraction, respectively. We first noticed that the estimated mixing proportions of both models were different from those of the joint development model without any covariates, and the proportions of these two models were also different. Upon further examination, the clusters obtained from the joint development model in Section \ref{app:number} agreed better with those from the model in Table \ref{tbl:GMM_selection} (i.e., the GMM with covariates from feature selection) (Dumenci's Latent Kappa\footnote{Note that the equations to calculate latent Kappa is the same as those to calculate Kappa statistics. The only difference lies in that the former was developed for the latent categorical variables \citep{Dumenci2011kappa, Dumenci2019knee}. In this project, we calculate latent Kappa using the \textit{R} package \textit{fmsb} \citep{Nakazawa2019fmsb}.}: $0.75$ with $95\%$ CI ($0.70$, $0.80$), $74$ out of $500$ students were assigned to different classes by the two models) than those from the model in Table \ref{tbl:GMM_reduction} (i.e., the GMM with covariates from feature reduction) (Kappa statistics: $0.69$ with $95\%$ CI ($0.63$, $0.74$), $98$ out of $500$ students were assigned to different classes by two models). The agreement of latent classes of the two GMM with covariates was better (Kappa statistics: $0.79$ with $95\%$ CI ($0.75$, $0.84$), $62$ out of $500$ students were assigned to different classes by the two models). Additionally, the clustering models can identify students in Class $3$ from the other two latent classes, although the students in Class $1$ and $2$ may switch membership when adding or changing covariates in this example.

\tablehere{5}

\tablehere{6}

We observed that the class-specific estimates obtained from the two GMMs for joint development are similar. On average, students in Class $1$ had the lowest reading and mathematics achievement levels during the entire study duration. On average, students in Class $2$ had a better academic performance at the beginning of the survey and during the pre-knot development stage than those in Class $1$. Students in Class $3$ had the best academic performance on average throughout the entire duration. Post-knot development in reading and mathematics skills slowed substantially for all three classes, and the change to the slower growth rate occurred earlier in reading ability than in mathematics ability. Additionally, students who had higher reading ability performance tended to have higher mathematics IRT scores in general. However, the association in the development of two abilities varied among the three classes. Specifically, this association was significant during the whole duration for students in Class $1$, while significant at the beginning and during the pre-knot development stage for students in Class $2$, and while significant only at the initial status for students in Class $3$. 

The effects of baseline characteristics were different in the two models. Specifically, from the model in Table \ref{tbl:GMM_selection}, students from families with higher income and parents' education were more likely to be in Class $2$, while girls from families with higher socioeconomic status were more likely to be in Class $3$. In addition to the above insights, the model in Table \ref{tbl:GMM_reduction} also suggests that students with higher values in the scores of factor $1$ (i.e., the differentiation in teacher-reported ability and teacher-rated problems) tended to have better academic performance (i.e., be separated in Class $2$ or $3$). 

\section{Discussion}\label{discussion}
This article proposes a GMM with a PBLSGM as the within-class model to investigate the heterogeneity in joint nonlinear development and the effects that baseline characteristics have on the class membership. We demonstrate the proposed model using simulation studies and a real-world data analysis. We performed in-depth investigations on the convergence rate, the performance metrics, and the accuracy value of the clustering algorithm through simulation studies. The convergence rate of the proposed model achieved at least $94\%$ under the conditions where two repeated outcomes were correlated. In general, we obtained unbiased and accurate point estimates with target coverage probabilities from the proposed GMM. The simulation study also showed that several factors, especially the separation between outcome-specific knot locations and the correlation between the two outcomes, affect the accuracy value of the clustering algorithm. The accuracy values were at least $80\%$ when we set the difference in outcome-specific knot locations as $1.00$. We also illustrate how to apply the proposed model on a random subset with $n=500$ from ECLS-K:2011 and provide general steps for implementing the model in practice. 

In addition to providing insights of the association between developmental processes and the heterogeneity in such associations, the proposed model can also improve performance measures and clustering performance (i.e., accuracy), as shown in the simulation study. This is not unexpected. As stated earlier, the Mahalanobis distance and the separation between knot locations increase when we consider multiple developmental processes simultaneously. The increased separation between two latent classes improves the clustering performance, quantified by the accuracy value; the higher accuracy value, indicating the algorithm groups more trajectories into the correct clusters, results in better within-class estimates that are only based on the trajectories separated into the corresponding cluster. We also noticed that the estimated mixing proportions obtained from the joint development model and each univariate development model differed in the simulation study and the real-world data analysis. For the empirical example, this difference can be attributed to the joint development model focusing on class membership on both reading and mathematics ability, whereas the univariate development model can only separate reading or mathematics ability trajectories. 

\subsection{Practical Considerations}
As shown in the simulation study, the proposed model suffered a severe non-convergence issue (the convergence rate was only $54\%$ under some conditions) when the two processes in the `joint' development were actually isolated (i.e., the between-construct growth factor covariance was $0$). Accordingly, whether to investigate the heterogeneity in joint development should be decided before constructing the proposed GMM. If it is of interest, we recommend testing whether the processes are associated using the PBLSGM that converged well even under conditions with zero between-construct correlation, as shown in \citet{Liu2021PBLSGM}. Additionally, the challenges of the GMM with univariate development, such as determining the optimal number of latent classes and deciding the covariates for the model, are even more challenging for the GMM with joint development. Given these considerations, we recommend building the proposed GMM in a stepwise fashion, although we proposed it as a one-step model. The recommended steps are:

\begin{enumerate}
\item{Univariate Analysis: Exclusion of any covariates, construct a latent growth curve model and growth mixture models with different numbers of latent classes to select the optimal number by the BIC for each developmental process,}
\item{Association Analysis: Test whether the processes are associated using the PBLSGM,}
\item{If at least one optimal number of latent classes from Step 1 is two or above and these processes are associated, then construct GMMs for joint development:}
\begin{enumerate}
\item{Decide the number of latent classes, which can be driven by the BIC as in Step 1 or driven by answering a question as we did in the Application section,}
\item{Conduct feature selection by constructing a SEM Forests model, or feature extraction by conducting the EFA, or both to shrink covariate space,}
\item{Build the proposed GMM(s) with covariate set(s) obtained from Step 3(b).}
\end{enumerate}
\end{enumerate}

In addition to the stepwise list, we still want to present a collection of recommendations for potential issues that empirical researchers may face in practice. First of all, as shown in the Application section, the number of latent classes of the joint development model decided by the BIC can differ from that of the univariate development models. This is not surprising as the addition of one latent class of the proposed model includes additional $32$ parameters; for such a complex model, the BIC tends to select the model with fewer latent classes as it penalizes model complexity (i.e., the number of parameters). Alternatively, the selection of the optimal number of clusters can be driven by \textit{a priori} knowledge obtained from univariate development models, as we did in the Application section. In practice, suppose we have already known that three latent classes exist in the development of reading or mathematics ability; one reasonable research question is to examine the heterogeneity in joint development of reading and mathematics skills with the three clusters assumption. 

Furthermore, we need to decide to include which covariates in the proposed model to inform the cluster formation. The covariate space usually has a high dimension and highly correlated subsets in educational and psychological domains where the GMM is widely used. Two approaches, feature selection and feature extraction, can help reduce covariate dimension and address the possible collinearity issue in the statistical and machine learning literature. Both have their counterparts in the SEM framework. For example, the SEM Forests can select covariates with the highest importance scores in predicting the model implied mean vector and variance-covariance structure, while the EFA can reduce the number of covariates by replacing a large number of those highly correlated variables with a small number of factors. Early studies have demonstrated that these two methods can be used for the GMM with univariate development. In this study, we extended these two methods to the GMM with joint development. 

Though it is not our aim to comprehensively examine both methods in this study, we still want to add four notes on these approaches for empirical researchers. First, no method is universally preferred. For the SEM Forest, variable importance scores are generated by the template growth curve model and candidate covariates. As shown in the Application section, we can construct a SEM Forests model for joint development and each univariate development, which allows us to evaluate the (relative) importance of covariates for each model, and then examine the causes of the heterogeneity in each development efficiently. For example, patients' highest education was the factor with a high importance score in three models. However, its impact on the developmental processes varied, as its relative importance score is much higher in reading development than in mathematics development. This difference can also provide insights to us: reading ability is an ability that is more related to exposure, and parents with a higher educational level can provide a better environment that helps improve reading skills. We cannot obtain such insights from the feature extraction method since the EFA produces factors that can explain the variance of a larger data set only from the covariates themselves.

Second, the variable importance scores of SEM Forests only tell us which characteristics have a (relatively) greater impact on sample heterogeneity, but it does not provide how these characteristics affect the heterogeneity. For example, sex had opposite effects on the univariate development of reading ability and mathematics ability. Specifically, boys tended to perform better in mathematics while girls outperformed in reading, which can only be observed by adding the covariate to the univariate development models. 

Additionally, the estimated mixing proportions varied if we included different covariates in the model for joint development. We have the same issue for GMMs with univariate development. Because we decided on the covariates by data-driven approaches and the agreement between the cluster labels obtained from the two models is good enough\footnote{As \citet{Nakazawa2019fmsb, Landis1977kappa}, a value of latent Kappa above $0.8$ indicates an almost perfect agreement, while a value greater than $0.6$ suggests a substantial agreement. }, it should not be a major concern to put which covariate set in the model; instead, a question of greater research interest could be examining the individuals who were re-classified by different models. For example, the proposed algorithm can identify students in Class $3$, but sometimes failed to distinguish students in the other two classes, which may suggest that the boundary between students in Class $1$ and $2$ is not so consistent when we add or change covariate(s). 

Last, the interpretation of logistic coefficients of the GMM with covariates obtained from the two methods could be different. For example, the GMM with covariates from the feature selection method only suggests that socioeconomic variables were positively associated with academic performance, while that with covariates from the feature extraction method also suggests that teacher-reported abilities and approach-to-learning were positively associated with academic performance while internalizing/externalizing problems were negatively associated with academic achievement. Again, it is not our aim to compare and contrast the two approaches to shrinking covariate space; instead, we want to demonstrate how to obtain a more holistic evaluation of the heterogeneity in joint development by the proposed model and these methods. 

\subsection{Methodological Considerations and Future Directions}
There are several future directions for the current study. First of all, we assumed that the outcome-specific knots in each latent class are roughly the same across individuals (i.e., the heterogeneity in knot locations is only due to the existence of subpopulations) to build a parsimonious model. However, the outcome-specific knots in each cluster can also be individually different. Accordingly, one possible future model is to relax the fixed class-specified knots assumption and examine their random effects so that we can assess individual-level knots.

Second, in this study, we build the model assuming that all baseline covariates only have indirect effects on the heterogeneity in trajectories (i.e., only informing the cluster formation). However, these covariates can also directly affect the heterogeneity in trajectories (i.e., explaining the variance of class-specific growth factors) \citep{Kim2016expert, Masyn2017Direct, Liu2020MoE}. The proposed model can also be extended accordingly. One limitation of this assumption lies in that it does not allow for examining time-varying covariates. Fortunately, \citet{Liu2020MoE} has demonstrated how to evaluate the effects of time-varying covariates on the heterogeneity in univariate development. It should be extended for analyzing joint development straightforwardly. 

Third, the mixing proportions in the current study are determined by the logistic functions that only allow for (generalized) linear models, which is also a possible explanation that the variables approach-to-learning and attentional focus were important in the SEM Forests but were not statistically significant in the GMM model. Accordingly, another possible extension of the current study is to develop other functional forms for the mixing proportions that may allow for nonlinear models. 

\subsection{Concluding Remarks}
In this article, we propose to use a GMM to investigate the heterogeneity in nonlinear joint development and the effects of the baseline characteristics on sample heterogeneity with an assumption that these characteristics only inform the cluster formation. Overall, we have shown the performance and application of the GMM with a PBLSGM as the within-class model. Given that the nonlinear underlying developmental process could be other functional forms other than bilinear spline for either theoretical or empirical considerations, we provide the GMM with joint quadratic trajectories and that with joint Jenss-Bayley growth curve in the online appendix for researchers who are willing to utilize them. 

\bibliographystyle{apalike}
\bibliography{Extension3}

\newpage
\renewcommand\thefigure{\arabic{figure}}
\setcounter{figure}{0}
\begin{figure}
\centering
\includegraphics[width=0.98\textwidth]{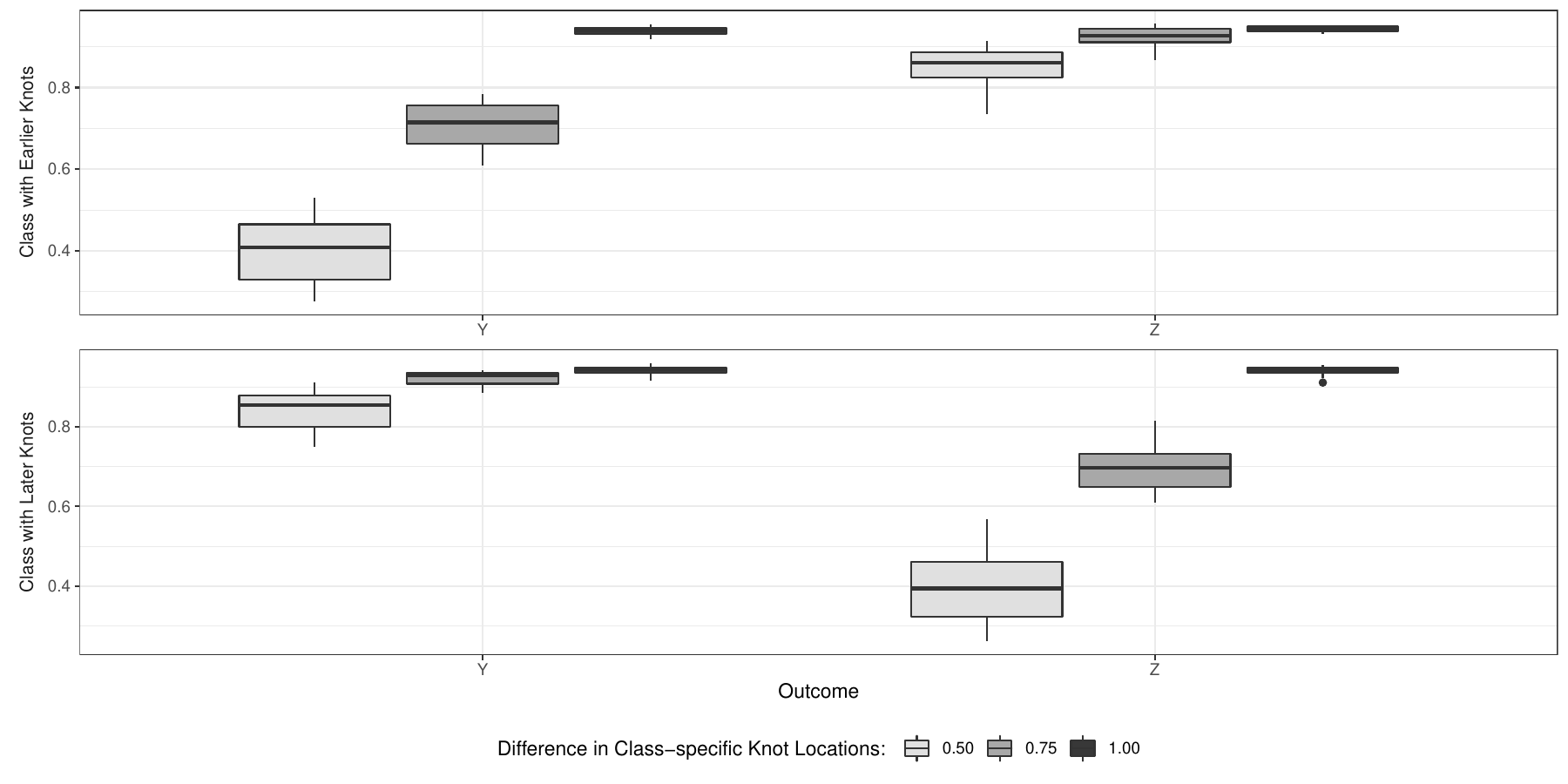}
\caption{Coverage Probabilities of Outcome-Specific Knot in Each Cluster}
\label{fig:KnotCP}
\end{figure}

\begin{figure}
\centering
\includegraphics[width=1.0\textwidth]{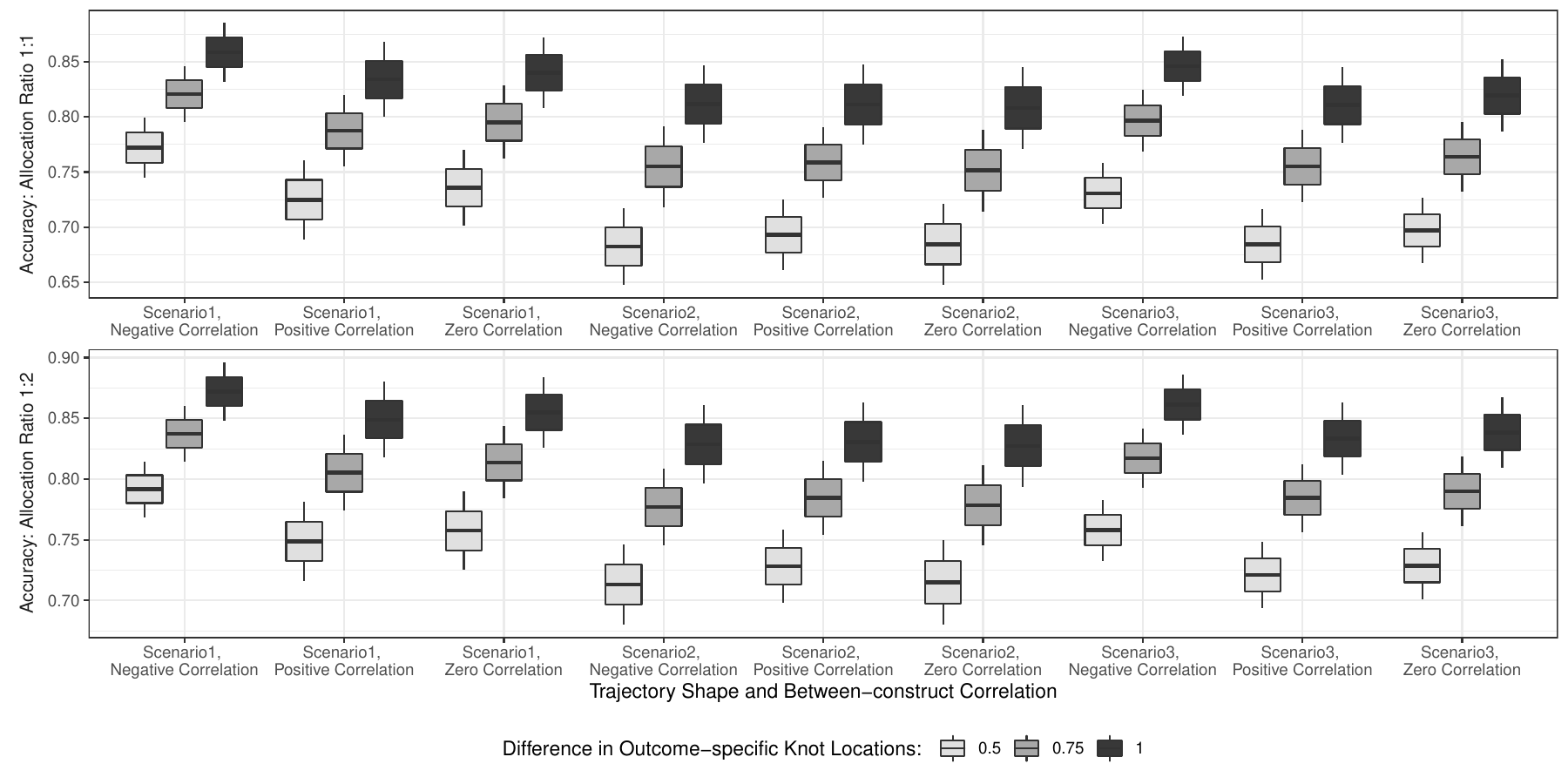}
\caption{Mean Accuracy of the Proposed Model across All Conditions}
\label{fig:acc}
\end{figure}

\begin{figure}
\centering
\includegraphics[width=1.0\textwidth]{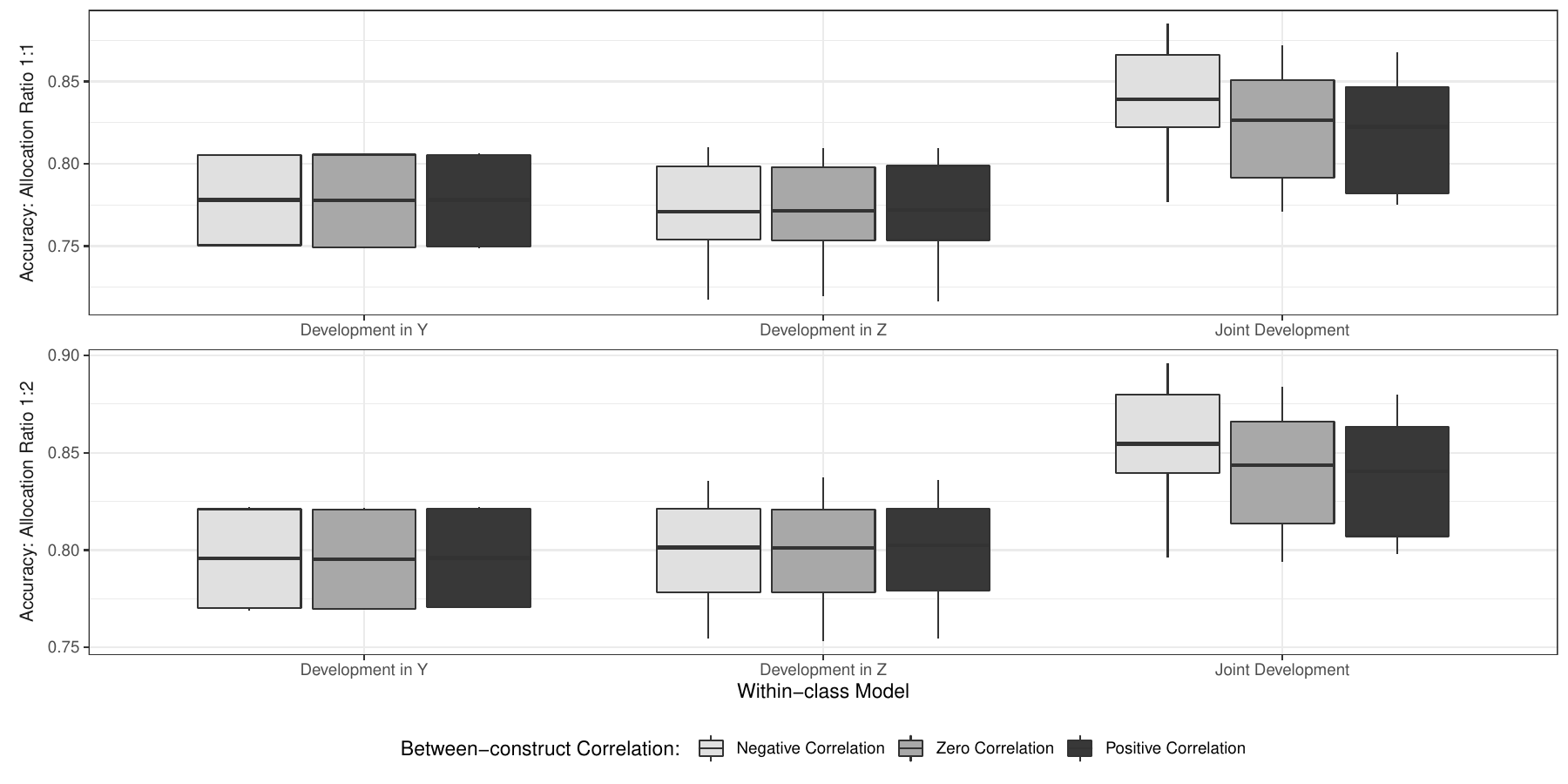}
\caption{Comparison of Mean Accuracy of GMMs for Joint Development and Univariate Development with Large Separation in Outcome-specific Knot Locations}
\label{fig:acc_compare}
\end{figure}

\begin{figure}
\begin{subfigure}{.50\textwidth}
\centering
\includegraphics[width=1.0\linewidth]{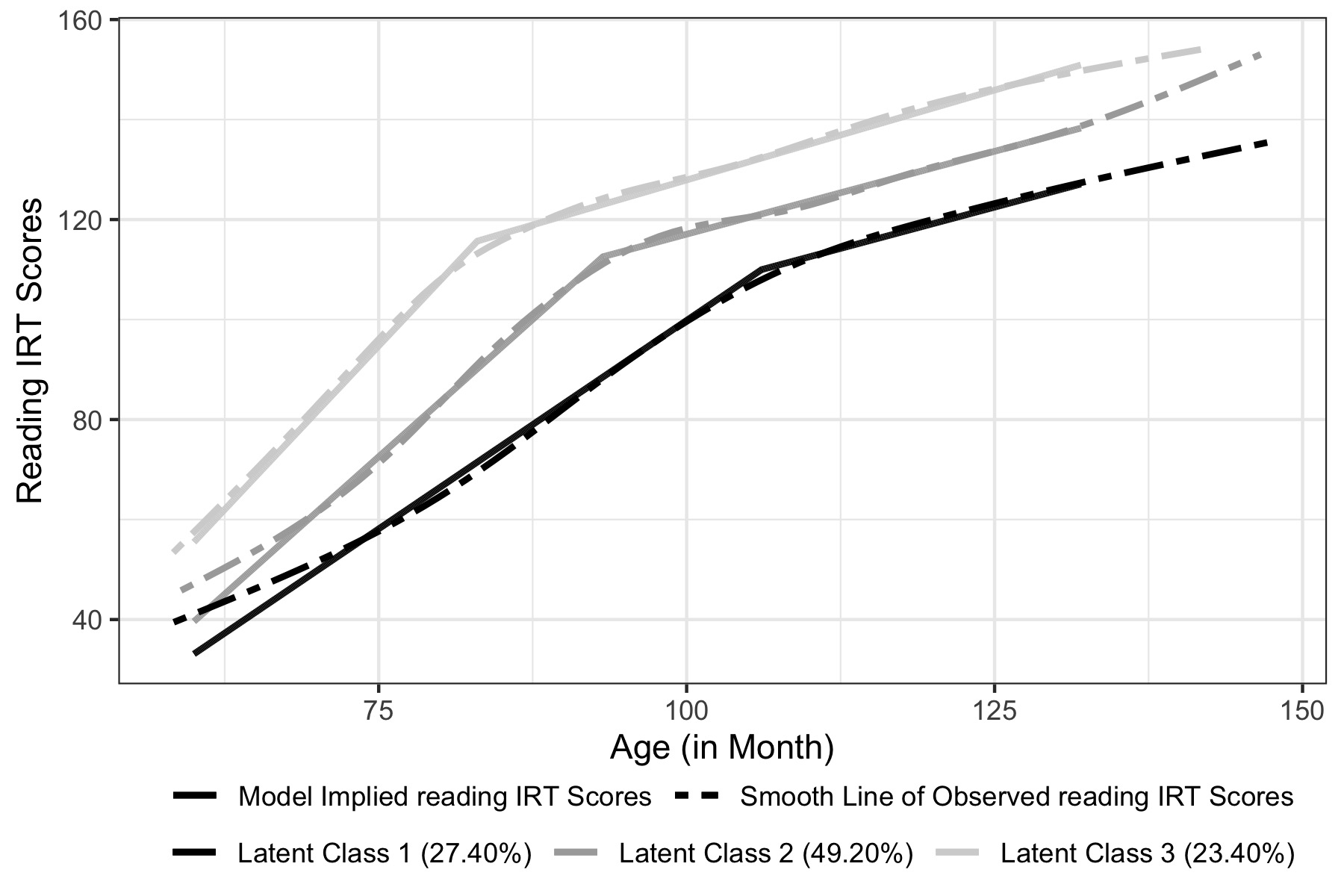}
\caption{Trajectory of Reading Ability}
\label{fig:traj_reading_uni}
\end{subfigure}%
\begin{subfigure}{.50\textwidth}
\centering
\includegraphics[width=1.0\linewidth]{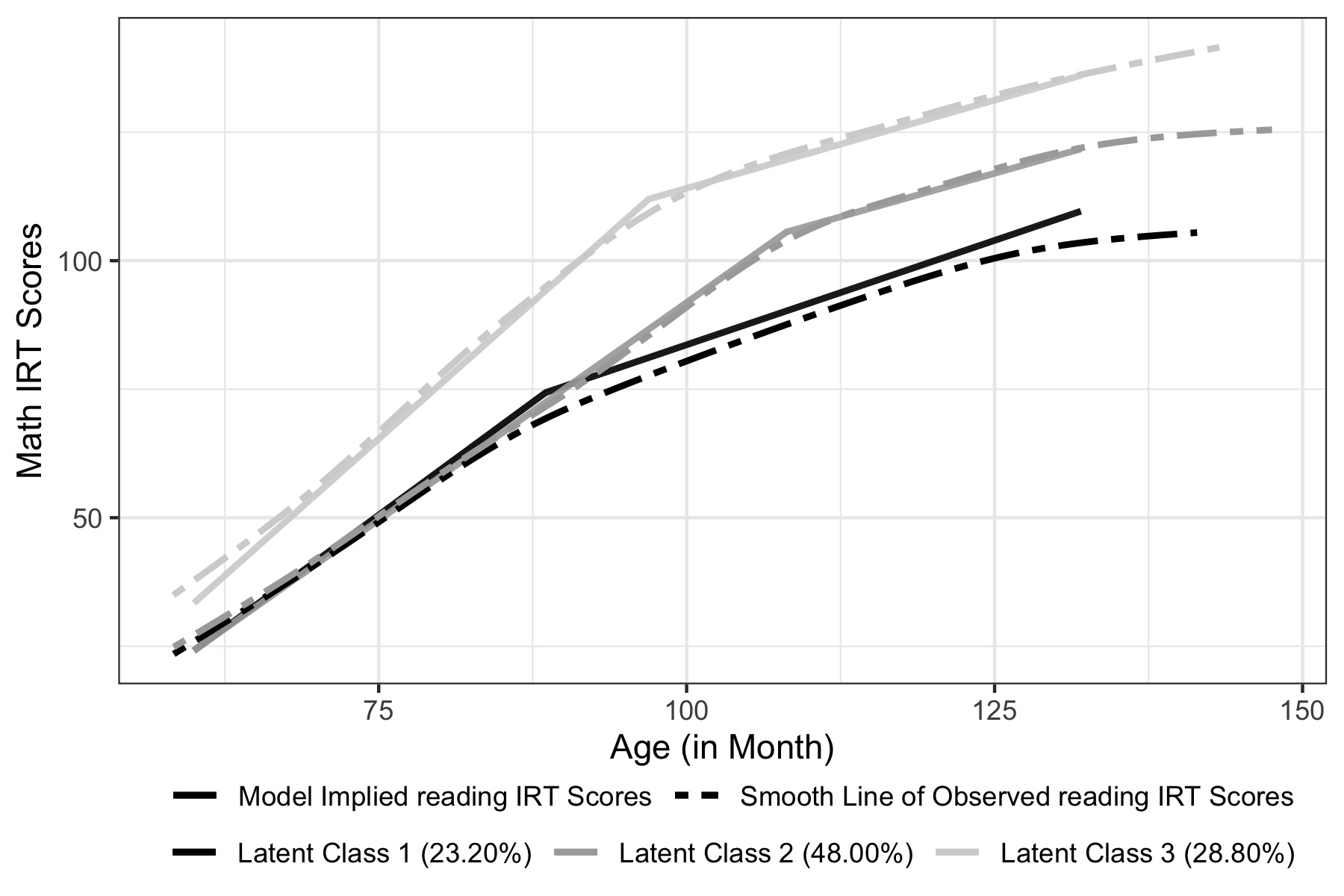}
\caption{Trajectory of Math Ability}
\label{fig:traj_math_uni}
\end{subfigure}
\caption{Model Implied Trajectory and Smooth Line of Univariate Repeated Outcome}
\label{fig:traj_uni}
\end{figure}

\begin{figure}
\begin{subfigure}{.50\textwidth}
\centering
\includegraphics[width=1.0\linewidth]{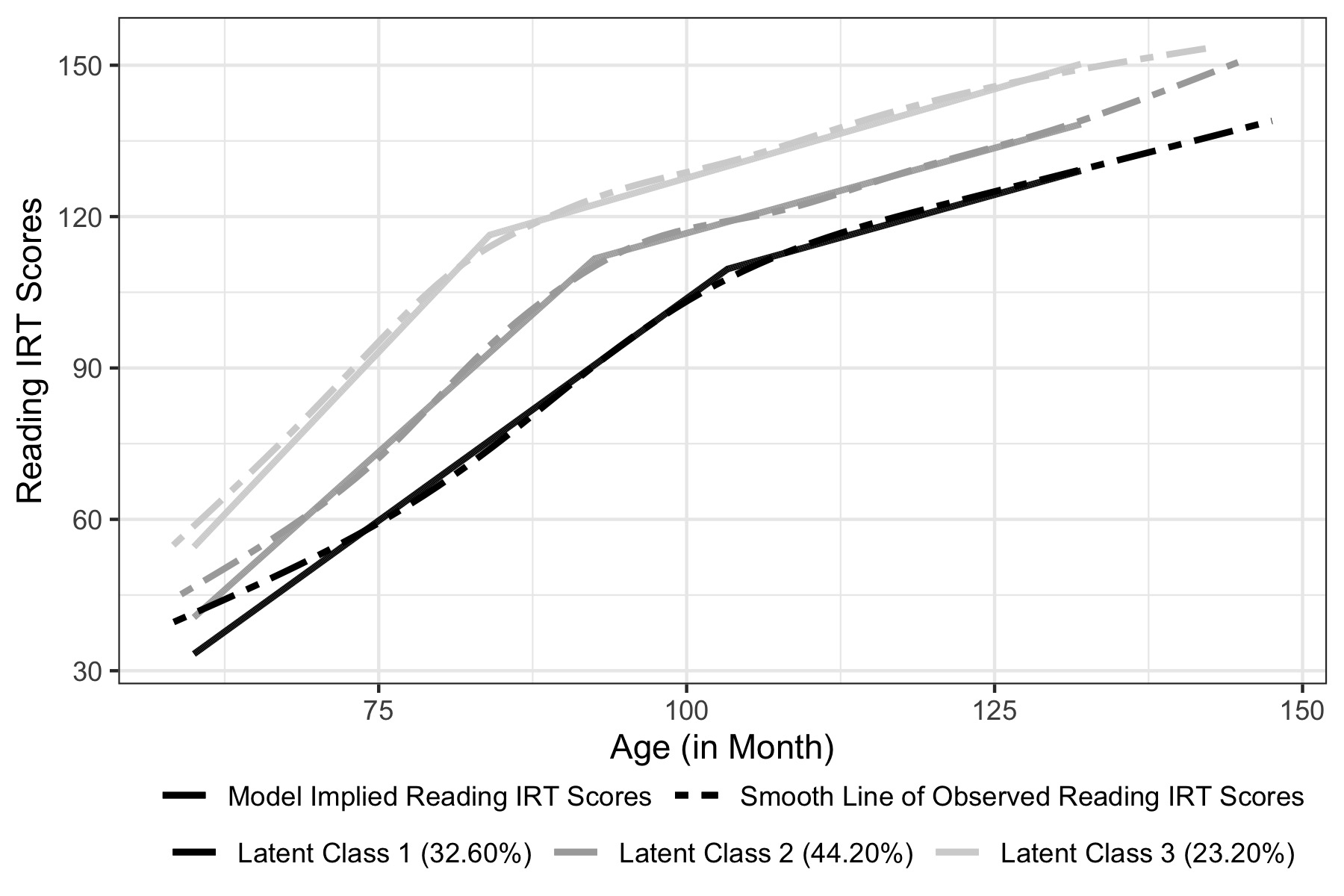}
\caption{Trajectory of Reading Ability}
\label{fig:traj_reading_bi}
\end{subfigure}%
\begin{subfigure}{.50\textwidth}
\centering
\includegraphics[width=1.0\linewidth]{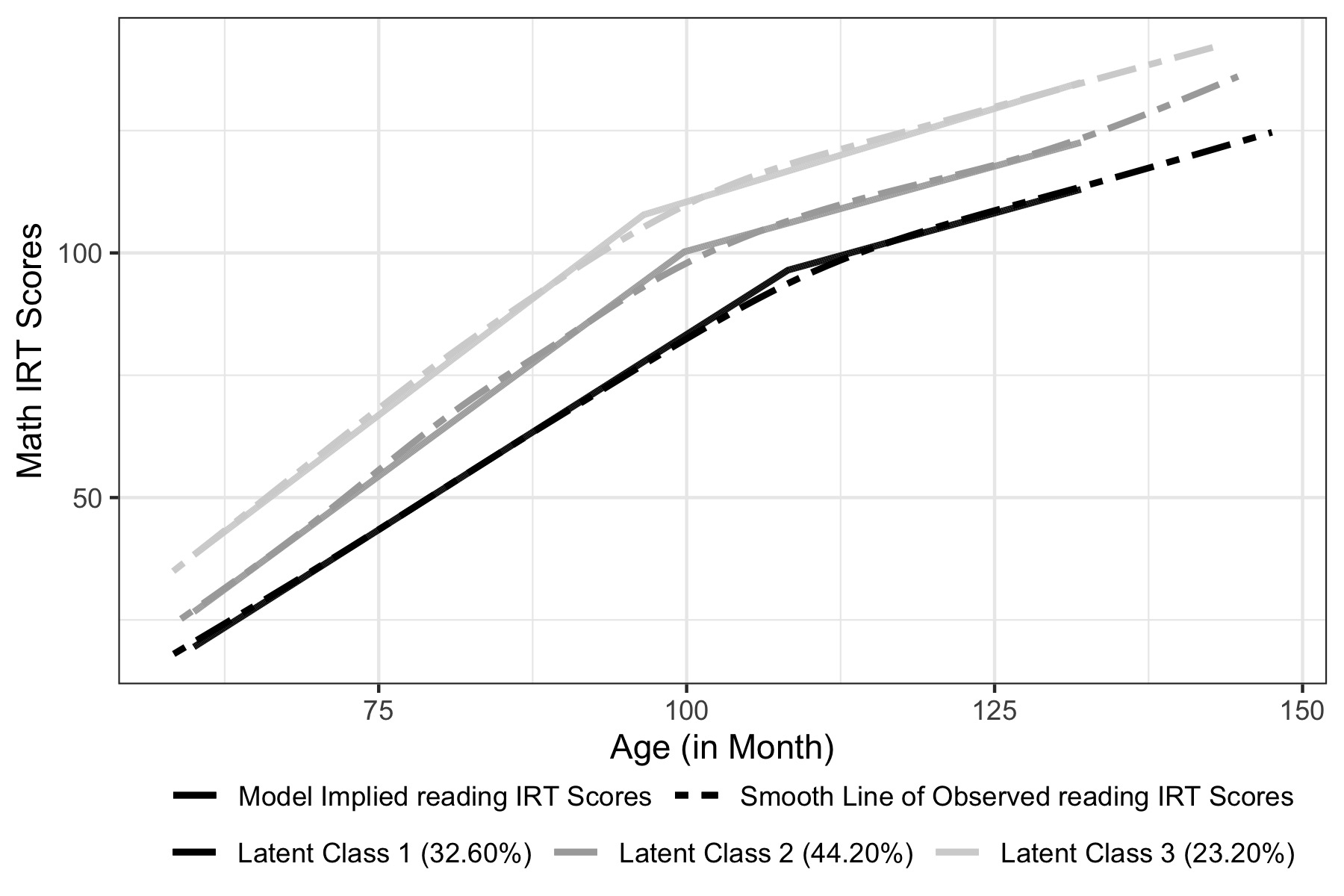}
\caption{Trajectory of Math Ability}
\label{fig:traj_math_bi}
\end{subfigure}
\caption{Model Implied Trajectory and Smooth Line of Bivariate Repeated Outcome}
\label{fig:traj_bi}
\end{figure}

\newpage
\renewcommand\thetable{\arabic{table}}
\setcounter{table}{0}

\begin{table}
\centering
\begin{threeparttable}
\setlength{\tabcolsep}{5pt}
\renewcommand{\arraystretch}{0.75}
\caption{Performance Metrics: Definitions and Estimates}
\begin{tabular}{p{4cm}p{4.5cm}p{5.5cm}}
\hline
\hline
\textbf{Criteria} & \textbf{Definition} & \textbf{Estimate} \\
\hline
Relative Bias & $E_{\hat{\theta}}(\hat{\theta}-\theta)/\theta$ & $\sum_{s=1}^{S}(\hat{\theta}_{s}-\theta)/S\theta$ \\
Empirical SE & $\sqrt{Var(\hat{\theta})}$ & $\sqrt{\sum_{s=1}^{S}(\hat{\theta}_{s}-\bar{\theta})^{2}/(S-1)}$ \\
Relative RMSE & $\sqrt{E_{\hat{\theta}}(\hat{\theta}-\theta)^{2}}/\theta$ & $\sqrt{\sum_{s=1}^{S}(\hat{\theta}_{s}-\theta)^{2}/S}/\theta$ \\
Coverage Probability & $Pr(\hat{\theta}_{\text{low}}\le\theta\le\hat{\theta}_{\text{upper}})$ & $\sum_{s=1}^{S}I(\hat{\theta}_{\text{low},s}\le\theta\le\hat{\theta}_{\text{upper},s})/S$\\
\hline
\hline
\end{tabular}
\label{tbl:metric}
\begin{tablenotes}
\small
\item[1]{$\theta$: the population value of the parameter of interest} \\
\item[2]{$\hat{\theta}$: the estimate of $\theta$} \\
\item[3]{$S$: the number of replications and set as $1,000$ in our simulation study} \\
\item[4]{$s=1,\dots,S$: indices the replications of the simulation} \\
\item[5]{$\hat{\theta}_{s}$: the estimate of $\theta$ from the $s^{th}$ replication} \\
\item[6]{$\bar{\theta}$: the mean of $\hat{\theta}_{s}$'s across replications} \\
\item[7]{$I()$: an indicator function}
\end{tablenotes}
\end{threeparttable}
\end{table}

\begin{table}
\centering
\resizebox{1.15\textwidth}{!}{
\begin{threeparttable}
\caption{Simulation Design for the Proposed GMM (Within-class Model: PBLSGM with Fixed Knots)}
\begin{tabular}{p{8cm}p{12.8cm}}
\hline
\hline
\multicolumn{2}{c}{\textbf{Fixed Conditions}} \\
\hline
\textbf{Variables} & \textbf{Conditions} \\
\hline
Variance of Intercept & $\psi_{00}^{(k)[u]}=25$, $u=y, z$; $k=1, 2$ \\
\hline
Variance of Slopes & $\psi_{11}^{(k)[u]}=\psi_{22}^{(k)[u]}=1$, $u=y, z$; $k=1, 2$\\
\hline
Correlations of GFs & $\rho^{(k)[u]}=0.3$, $u=y, z$; $k=1, 2$ \\
\hline
Time (\textit{t}) & $10$ scaled and equally spaced $t_{j} (j=0, \cdots, J-1, J=10)$ \\
\hline
Individual \textit{t} & $t_{ij} \sim U(t_{j}-\Delta, t_{j}+\Delta) (j=0, \cdots, J-1; \Delta=0.25)$ \\
\hline
Sample Size & $n=500$ \\
\hline
Within-construct Mahalanobis distance & $d=0.86$ \\
\hline
Residual Correlation & $\rho_{\epsilon}=0.3$ \\
\hline
\hline
\multicolumn{2}{c}{\textbf{Manipulated Conditions}} \\
\hline
\hline
\textbf{Variables} & \textbf{Conditions} \\
\hline
\multirow{2}{*}{Logistic Coefficients} & $\beta_{0}=0$ (Allocation ratio is about $1:1$), $\beta_{1}=\log(1.5)$, $\beta_{2}=\log(1.7)$\\
& $\beta_{0}=0.775$ (Allocation ratio is about $1:2$), $\beta_{1}=\log(1.5)$, $\beta_{2}=\log(1.7)$ \\
\hline
Residual Variance & $\theta_{\epsilon}^{(k)[u]}=1$ or $2$, $u=y, z$; $k=1, 2$ \\
\hline
\multirow{3}{*}{Locations of knots} & $\mu_{\gamma}^{(1)[y]}=4.00$; $\mu_{\gamma}^{(2)[y]}=4.50$; $\mu_{\gamma}^{(1)[z]}=4.50$; $\mu_{\gamma}^{(2)[z]}=5.00$ \\
& $\mu_{\gamma}^{(1)[y]}=3.75$; $\mu_{\gamma}^{(2)[y]}=4.50$; $\mu_{\gamma}^{(1)[z]}=4.50$; $\mu_{\gamma}^{(2)[z]}=5.25$ \\
& $\mu_{\gamma}^{(1)[y]}=3.50$; $\mu_{\gamma}^{(2)[y]}=4.50$; $\mu_{\gamma}^{(1)[z]}=4.50$; $\mu_{\gamma}^{(2)[z]}=5.50$ \\
\hline
Between-construct Correlation of GF & $\rho=-0.3,0,0.3$ \\
\hline
\hline
\multicolumn{2}{l}{\textbf{Scenario 1: Different Intercept Mean and Knot Mean for $u=y$ and $u=z$}} \\
\hline
\textbf{Variables} & \textbf{Conditions} \\
\hline
Means of Slope 1's & $\mu_{\eta_{1}}^{(k)[u]}=5$ $(k=1, 2)$ \\
\hline
Means of Slope 2's
& $\mu_{\eta_{2}}^{(k)[u]}=2.6$ $(k=1, 2)$ \\
\hline
Means of Intercepts & $\mu_{\eta_{0}}^{(1)[y]}=98$, $\mu_{\eta_{0}}^{(2)[y]}=102$, $\mu_{\eta_{0}}^{(1)[z]}=98$, $\mu_{\eta_{0}}^{(2)[z]}=102$ \\
\hline
\hline
\multicolumn{2}{l}{\textbf{\makecell[l]{Scenario 2: Different Intercept Mean and Knot Mean for $u=y$, \\ Different First Slope Mean and Knot Mean for $u=z$}}}\\
\hline
\textbf{Variables} & \textbf{Conditions} \\
\hline
Means of Intercepts & $\mu_{\eta_{0}}^{(1)[y]}=98$, $\mu_{\eta_{0}}^{(2)[y]}=102$, $\mu_{\eta_{0}}^{(k)[z]}=100$ $(k=1,2)$ \\
\hline
Means of Slope 2's & $\mu_{\eta_{2}}^{(k)[y]}=2.6$, $\mu_{\eta_{2}}^{(k)[z]}=2$ \\
\hline
Means of Slope 1's & $\mu_{\eta_{1}}^{(k)[y]}=5.0$ $(k=1,2)$, $\mu_{\eta_{1}}^{(1)[z]}=4.4$, $\mu_{\eta_{1}}^{(2)[z]}=3.6$ \\
\hline
\hline
\multicolumn{2}{l}{\textbf{\makecell[l]{Scenario 3: Different Intercept Mean and Knot Mean for $u=y$, \\ Different Second Slope Mean and Knot Mean for $u=z$}}}\\
\hline
\textbf{Variables} & \textbf{Conditions} \\
\hline
Means of Intercepts & $\mu_{\eta_{0}}^{(1)[y]}=98$, $\mu_{\eta_{0}}^{(2)[y]}=102$, $\mu_{\eta_{0}}^{(k)[z]}=100$ $(k=1,2)$ \\
\hline
Means of Slope 1's & $\mu_{\eta_{1}}^{(k)[y]}=5.0$, $\mu_{\eta_{1}}^{(k)[z]}=4.4$ \\
\hline
Means of Slope 2's & $\mu_{\eta_{2}}^{(k)[y]}=2.6$ $(k=1,2)$, $\mu_{\eta_{2}}^{(1)[z]}=2.0$, $\mu_{\eta_{2}}^{(2)[z]}=2.8$ \\
\hline
\hline
\end{tabular}
\label{tbl:simu}
\end{threeparttable}}
\end{table}

\begin{table}
\centering
\resizebox{1.15\textwidth}{!}{
\begin{threeparttable}
\caption{Summary of Model Fit Information for GMMs}
\begin{tabular}{lrrrrrrr}
\hline
\hline
\multicolumn{7}{c}{\textbf{GMM with Development in Reading Ability}}\\
\hline
$\#$ of Clusters & -2ll & AIC & BIC & $\%$ of Class $1$ & $\%$ of Class $2$ & $\%$ of Class $3$ & $\#$ of Para. \\
\hline
$1$ & $32696.90$ & $32718.90$ &$32765.27$ & $100.0\%$ & ---\tnote{1} & --- & $11$\\
$2$ & $32298.79$ & $32344.79$ &$32441.73$ & $33.4\%$ & $66.6\%$ & --- & $23$ \\
$3$& $32133.83$ & $32203.83$ & $32351.34$ & $27.4\%$ & $49.2\%$ & $23.4\%$ & $35$ \\
\hline
\hline
\multicolumn{7}{c}{\textbf{GMM with Development in Mathematics Ability}}\\
\hline
$\#$ of Clusters & -2ll & AIC & BIC & $\%$ of Class $1$ & $\%$ of Class $2$ & $\%$ of Class $3$ & $\#$ of Para. \\
\hline
$1$ & $31579.12$ & $31601.12$ &$31647.48$ & $100.0\%$ & --- & --- & $11$\\
$2$ & $31385.75$ & $31431.75$ & $31528.68$ & $36.6\%$ & $63.4\%$ & --- & $23$\\
$3$& $31235.90$ & $31305.90$ & $31453.42$ & $23.2\%$ & $48.0\%$ & $28.8\%$ & $35$\\
\hline
\hline
\multicolumn{7}{c}{\textbf{GMM with Joint Development in Reading \& Mathematics Ability}}\\
\hline
$\#$ of Clusters & -2ll & AIC & BIC & $\%$ of Class $1$ & $\%$ of Class $2$ & $\%$ of Class $3$ & $\#$ of Para. \\
\hline
$1$ & $63328.02$ & $63392.02$ & $63526.89$ & $100.0\%$ & --- & --- & $32$\\
$2$ & $62793.03$ & $62923.03$ & $63268.39$ & $34.0\%$ & $66.0\%$ & --- & $65$\\
$3$& $62573.31$ & $62769.31$ & $63318.20$ & $32.6\%$ & $44.2\%$ & $23.2\%$ & $98$\\
\hline
\hline
\end{tabular}
\label{tbl:compare}
\begin{tablenotes}
\small
\item[2] --- indicates that the metric was not available for the model.
\end{tablenotes}
\end{threeparttable}}
\end{table}

\begin{table}
\centering
\resizebox{1.15\textwidth}{!}{
\begin{threeparttable}
\setlength{\tabcolsep}{5pt}
\renewcommand{\arraystretch}{0.75}
\caption{Estimates of GMM with Joint Development of Reading and Mathematics Ability with Covariates from Feature Selection}
\begin{tabular}{lrrrrrr}
\hline
\hline
& \multicolumn{6}{c}{\textbf{Reading Ability}} \\
\hline
& \multicolumn{2}{c}{\textbf{Class 1 ($\boldsymbol{30.6\%}$)}} & \multicolumn{2}{c}{\textbf{Class 2 ($\boldsymbol{50.2\%}$)}}& \multicolumn{2}{c}{\textbf{Class 3 ($\boldsymbol{19.2\%}$)}} \\
\hline
\textbf{Mean of Growth Factor} & Estimate (SE) & P value & Estimate (SE) & P value & Estimate (SE) & P value \\
\hline
\textbf{Intercept}\tnote{1} & $33.478$ ($0.932$) & $<0.0001^{\ast}$ & $39.998$ ($0.878$) & $<0.0001^{\ast}$ & $55.155$ ($2.140$) & $<0.0001^{\ast}$ \\
\textbf{Slope $1$} & $1.697$ ($0.047$) & $<0.0001^{\ast}$ & $2.197$ ($0.038$) & $<0.0001^{\ast}$ & $2.549$ ($0.085$) & $<0.0001^{\ast}$ \\
\textbf{Slope $2$} & $0.667$ ($0.040$) & $<0.0001^{\ast}$ & $0.678$ ($0.022$) & $<0.0001^{\ast}$ & $0.705$ ($0.022$) & $<0.0001^{\ast}$ \\
\hline
\hline
\textbf{Additional Parameter} & Estimate (SE) & P value & Estimate (SE) & P value & Estimate (SE) & P value \\
\hline
\textbf{Knot} & $104.404$ ($0.883$) & $<0.0001^{\ast}$ & $92.757$ ($0.456$) & $<0.0001^{\ast}$ & $84.475$ ($0.528$) & $<0.0001^{\ast}$ \\
\hline
\hline
\textbf{Variance of Growth Factor} & Estimate (SE) & P value & Estimate (SE) & P value & Estimate (SE) & P value \\
\hline
\textbf{Intercept} & $53.709$ ($13.141$) & $<0.0001^{\ast}$ & $63.992$ ($11.869$) & $<0.0001^{\ast}$ & $351.823$ ($61.995$) & $<0.0001^{\ast}$ \\
\textbf{Slope $1$} & $0.149$ ($0.024$) & $<0.0001^{\ast}$ & $0.132$ ($0.021$) & $<0.0001^{\ast}$ & $0.432$ ($0.094$) & $<0.0001^{\ast}$ \\
\textbf{Slope $2$} & $0.093$ ($0.021$) & $<0.0001^{\ast}$ & $0.034$ ($0.007$) & $<0.0001^{\ast}$ & $0.013$ ($0.006$) & $0.0303^{\ast}$ \\
\hline
\hline
& \multicolumn{6}{c}{\textbf{Mathematics Ability}} \\
\hline
& \multicolumn{2}{c}{\textbf{Class 1 ($\boldsymbol{30.6\%}$)}} & \multicolumn{2}{c}{\textbf{Class 2 ($\boldsymbol{50.2\%}$)}}& \multicolumn{2}{c}{\textbf{Class 3 ($\boldsymbol{19.2\%}$)}} \\
\hline
\textbf{Mean of Growth Factor} & Estimate (SE) & P value & Estimate (SE) & P value & Estimate (SE) & P value \\
\hline
\textbf{Intercept} & $19.128$ ($0.898$) & $<0.0001^{\ast}$ & $26.253$ ($0.776$) & $<0.0001^{\ast}$ & $38.773$ ($1.263$) & $<0.0001^{\ast}$ \\
\textbf{Slope $1$} & $1.572$ ($0.035$) & $<0.0001^{\ast}$ & $1.847$ ($0.029$) & $<0.0001^{\ast}$ & $1.909$ ($0.046$) & $<0.0001^{\ast}$ \\
\textbf{Slope $2$} & $0.695$ ($0.039$) & $<0.0001^{\ast}$ & $0.704$ ($0.024$) & $<0.0001^{\ast}$ & $0.761$ ($0.033$) & $<0.0001^{\ast}$ \\
\hline
\hline
\textbf{Additional Parameter} & Estimate (SE) & P value & Estimate (SE) & P value & Estimate (SE) & P value \\
\hline
\textbf{Knot} & $107.954$ ($1.016$) & $<0.0001^{\ast}$ & $100.232$ ($0.591$) & $<0.0001^{\ast}$ & $96.487$ ($0.696$) & $<0.0001^{\ast}$ \\
\hline
\hline
\textbf{Variance of Growth Factor} & Estimate (SE) & P value & Estimate (SE) & P value & Estimate (SE) & P value \\
\hline
\textbf{Intercept} & $46.387$ ($10.604$) & $<0.0001^{\ast}$ & $47.648$ ($9.781$) & $<0.0001^{\ast}$ & $107.252$ ($19.352$) & $<0.0001^{\ast}$ \\
\textbf{Slope $1$} & $0.067$ ($0.013$) & $<0.0001^{\ast}$ & $0.069$ ($0.012$) & $<0.0001^{\ast}$ & $0.073$ ($0.021$) & $0.0005^{\ast}$ \\
\textbf{Slope $2$} & $0.039$ ($0.015$) & $0.0093^{\ast}$ & $0.009$ ($0.009$) & $0.3173$ & $0.041$ ($0.012$) & $0.0006^{\ast}$ \\
\hline
\hline
& \multicolumn{6}{c}{\textbf{Association between Reading and Mathematics Ability}} \\
\hline
& \multicolumn{2}{c}{\textbf{Class 1 ($\boldsymbol{30.6\%}$)}} & \multicolumn{2}{c}{\textbf{Class 2 ($\boldsymbol{50.2\%}$)}}& \multicolumn{2}{c}{\textbf{Class 3 ($\boldsymbol{19.2\%}$)}} \\
\hline
\textbf{Mean of Growth Factor} & Estimate (SE) & P value & Estimate (SE) & P value & Estimate (SE) & P value \\
\hline
\textbf{Intercept} & $35.308$ ($9.365$) & $0.0002^{\ast}$ & $51.965$ ($8.804$) & $<0.0001^{\ast}$ & $75.535$ ($27.023$) & $0.0052^{\ast}$ \\
\textbf{Slope $1$} & $0.045$ ($0.014$) & $0.0013^{\ast}$ & $0.046$ ($0.012$) & $0.0001^{\ast}$ & $0.021$ ($0.032$) & $0.5117$ \\
\textbf{Slope $2$} & $0.026$ ($0.013$) & $0.0455^{\ast}$ & $-0.003$ ($0.005$) & $0.5485$ & $0.006$ ($0.007$) & $0.3914$ \\
\hline
\hline
& \multicolumn{6}{c}{\textbf{Logistic Coefficients}} \\
\hline
& \multicolumn{2}{c}{\textbf{Class 1 ($\boldsymbol{30.6\%}$)}} & \multicolumn{2}{c}{\textbf{Class 2 ($\boldsymbol{50.2\%}$)}}& \multicolumn{2}{c}{\textbf{Class 3 ($\boldsymbol{19.2\%}$)}} \\
\hline
& \multicolumn{2}{r}{OR ($95\%$ CI)\tnote{3}} & \multicolumn{2}{r}{OR ($95\%$ CI)} & \multicolumn{2}{r}{OR ($95\%$ CI)} \\
\hline
\textbf{Family Income} & \multicolumn{2}{r}{---\tnote{4}} & \multicolumn{2}{r}{$1.091$ ($1.020$, $1.165$)$^{\ast}$} & \multicolumn{2}{r}{$1.131$ ($1.043$, $1.227$)$^{\ast}$} \\
\textbf{Parents' Highest Education} & \multicolumn{2}{r}{---} & \multicolumn{2}{r}{$1.201$ ($1.006$, $1.434$)$^{\ast}$} & \multicolumn{2}{r}{$1.772$ ($1.410$, $2.226$)$^{\ast}$} \\
\textbf{Attentional Focus} & \multicolumn{2}{r}{---} & \multicolumn{2}{r}{$0.965$ ($0.678$, $1.373$)} & \multicolumn{2}{r}{$1.350$ ($0.907$, $2.010$)} \\
\textbf{approach-to-learning} & \multicolumn{2}{r}{---} & \multicolumn{2}{r}{$1.708$ ($0.868$, $3.362$)} & \multicolumn{2}{r}{$1.995$ ($0.919$, $4.329$)} \\
\textbf{Sex} ($0$---Boy; $1$---Girl) & \multicolumn{2}{r}{---} & \multicolumn{2}{r}{$1.217$ ($0.691$, $2.144$)} & \multicolumn{2}{r}{$2.227$ ($1.142$, $4.342$)$^{\ast}$} \\
\textbf{Race} ($0$---White; $1$---Others) & \multicolumn{2}{r}{---} & \multicolumn{2}{r}{$1.082$ ($0.567$, $2.066$)} & \multicolumn{2}{r}{$1.642$ ($0.792$, $3.407$)} \\
\hline
\hline
\end{tabular}
\label{tbl:GMM_selection}
\begin{tablenotes}
\small
\item[1] Intercept was defined as mathematics IRT scores at 60-month old in this case.
\item[2] $^{\ast}$ indicates statistical significance at $0.05$ level.
\item[3] OR ($95\%$ CI) indicates Odds Ratio ($95\%$ Confidence Interval).
\item[4] We set Class $1$ as the reference group.
\end{tablenotes}
\end{threeparttable}}
\end{table}

\begin{table}
\centering
\resizebox{1.15\textwidth}{!}{
\begin{threeparttable}
\setlength{\tabcolsep}{5pt}
\renewcommand{\arraystretch}{0.75}
\caption{Estimates of GMM with Joint Development of Reading and Mathematics Ability with Covariates from Feature Reduction}
\begin{tabular}{lrrrrrr}
\hline
\hline
& \multicolumn{6}{c}{\textbf{Reading Ability}} \\
\hline
& \multicolumn{2}{c}{\textbf{Class 1 ($\boldsymbol{20.0\%}$)}} & \multicolumn{2}{c}{\textbf{Class 2 ($\boldsymbol{59.8\%}$)}}& \multicolumn{2}{c}{\textbf{Class 3 ($\boldsymbol{20.2\%}$)}} \\
\hline
\textbf{Mean of Growth Factor} & Estimate (SE) & P value & Estimate (SE) & P value & Estimate (SE) & P value \\
\hline
\textbf{Intercept}\tnote{1} & $33.917$ ($1.093$) & $<0.0001^{\ast}$ & $38.670$ ($0.828$) & $<0.0001^{\ast}$ & $53.774$ ($1.979$) & $<0.0001^{\ast}$ \\
\textbf{Slope $1$} & $1.574$ ($0.052$) & $<0.0001^{\ast}$ & $2.159$ ($0.039$) & $<0.0001^{\ast}$ & $2.557$ ($0.077$) & $<0.0001^{\ast}$ \\
\textbf{Slope $2$} & $0.652$ ($0.055$) & $<0.0001^{\ast}$ & $0.657$ ($0.021$) & $<0.0001^{\ast}$ & $0.706$ ($0.020$) & $<0.0001^{\ast}$ \\
\hline
\hline
\textbf{Additional Parameter} & Estimate (SE) & P value & Estimate (SE) & P value & Estimate (SE) & P value \\
\hline
\textbf{Knot} & $108.144$ ($1.304$) & $<0.0001^{\ast}$ & $94.206$ ($0.513$) & $<0.0001^{\ast}$ & $84.837$ ($0.463$) & $<0.0001^{\ast}$ \\
\hline
\hline
\textbf{Variance of Growth Factor} & Estimate (SE) & P value & Estimate (SE) & P value & Estimate (SE) & P value \\
\hline
\textbf{Intercept} & $58.917$ ($14.522$) & $<0.0001^{\ast}$ & $72.138$ ($12.242$) & $<0.0001^{\ast}$ & $342.166$ ($54.118$) & $<0.0001^{\ast}$ \\
\textbf{Slope $1$} & $0.115$ ($0.024$) & $<0.0001^{\ast}$ & $0.128$ ($0.024$) & $<0.0001^{\ast}$ & $0.401$ ($0.081$) & $<0.0001^{\ast}$ \\
\textbf{Slope $2$} & $0.090$ ($0.030$) & $0.0027^{\ast}$ & $0.035$ ($0.007$) & $<0.0001^{\ast}$ & $0.012$ ($0.006$) & $0.0455^{\ast}$ \\
\hline
\hline
& \multicolumn{6}{c}{\textbf{Mathematics Ability}} \\
\hline
& \multicolumn{2}{c}{\textbf{Class 1 ($\boldsymbol{20.0\%}$)}} & \multicolumn{2}{c}{\textbf{Class 2 ($\boldsymbol{59.8\%}$)}}& \multicolumn{2}{c}{\textbf{Class 3 ($\boldsymbol{20.2\%}$)}} \\
\hline
\textbf{Mean of Growth Factor} & Estimate (SE) & P value & Estimate (SE) & P value & Estimate (SE) & P value \\
\hline
\textbf{Intercept} & $18.744$ ($1.104$) & $<0.0001^{\ast}$ & $25.097$ ($0.717$) & $<0.0001^{\ast}$ & $38.167$ ($1.151$) & $<0.0001^{\ast}$ \\
\textbf{Slope $1$} & $1.530$ ($0.040$) & $<0.0001^{\ast}$ & $1.829$ ($0.024$) & $<0.0001^{\ast}$ & $1.892$ ($0.036$) & $<0.0001^{\ast}$ \\
\textbf{Slope $2$} & $0.642$ ($0.051$) & $<0.0001^{\ast}$ & $0.696$ ($0.021$) & $<0.0001^{\ast}$ & $0.768$ ($0.030$) & $<0.0001^{\ast}$ \\
\hline
\hline
\textbf{Additional Parameter} & Estimate (SE) & P value & Estimate (SE) & P value & Estimate (SE) & P value \\
\hline
\textbf{Knot} & $110.725$ ($1.218$) & $<0.0001^{\ast}$ & $100.93$ ($0.043$) & $<0.0001^{\ast}$ & $96.505$ ($0.642$) & $<0.0001^{\ast}$ \\
\hline
\hline
\textbf{Variance of Growth Factor} & Estimate (SE) & P value & Estimate (SE) & P value & Estimate (SE) & P value \\
\hline
\textbf{Intercept} & $48.441$ ($12.058$) & $0.0001^{\ast}$ & $46.792$ ($8.952$) & $<0.0001^{\ast}$ & $103.326$ ($17.628$) & $<0.0001^{\ast}$ \\
\textbf{Slope $1$} & $0.063$ ($0.014$) & $<0.0001^{\ast}$ & $0.069$ ($0.012$) & $<0.0001^{\ast}$ & $0.075$ ($0.016$) & $<0.0001^{\ast}$ \\
\textbf{Slope $2$} & $0.042$ ($0.020$) & $0.0357^{\ast}$ & $0.013$ ($0.007$) & $0.0633$ & $0.038$ ($0.012$) & $0.0015^{\ast}$ \\
\hline
\hline
& \multicolumn{6}{c}{\textbf{Association between Reading and Mathematics Ability}} \\
\hline
& \multicolumn{2}{c}{\textbf{Class 1 ($\boldsymbol{20.0\%}$)}} & \multicolumn{2}{c}{\textbf{Class 2 ($\boldsymbol{59.8\%}$)}}& \multicolumn{2}{c}{\textbf{Class 3 ($\boldsymbol{20.2\%}$)}} \\
\hline
\textbf{Mean of Growth Factor} & Estimate (SE) & P value & Estimate (SE) & P value & Estimate (SE) & P value \\
\hline
\textbf{Intercept} & $36.751$ ($11.744$) & $0.0018^{\ast}$ & $55.778$ ($8.866$) & $<0.0001^{\ast}$ & $80.666$ ($23.683$) & $0.0007^{\ast}$ \\
\textbf{Slope $1$} & $0.038$ ($0.013$) & $0.0035^{\ast}$ & $0.047$ ($0.014$) & $0.0008^{\ast}$ & $0.021$ ($0.025$) & $0.4009$ \\
\textbf{Slope $2$} & $0.025$ ($0.020$) & $0.2113$ & $0.001$ ($0.005$) & $0.8415$ & $0.005$ ($0.008$) & $0.5320$ \\
\hline
\hline
& \multicolumn{6}{c}{\textbf{Logistic Coefficients}} \\
\hline
& \multicolumn{2}{c}{\textbf{Class 1 ($\boldsymbol{20.0\%}$)}} & \multicolumn{2}{c}{\textbf{Class 2 ($\boldsymbol{59.8\%}$)}}& \multicolumn{2}{c}{\textbf{Class 3 ($\boldsymbol{20.2\%}$)}} \\
\hline
& \multicolumn{2}{r}{OR ($95\%$ CI)\tnote{3}} & \multicolumn{2}{r}{OR ($95\%$ CI)} & \multicolumn{2}{r}{OR ($95\%$ CI)} \\
\hline
\textbf{Factor $1$} & \multicolumn{2}{r}{---\tnote{4}} & \multicolumn{2}{r}{$1.637$ ($1.019$, $2.631$)$^{\ast}$} & \multicolumn{2}{r}{$2.600$ ($1.403$, $4.818$)$^{\ast}$} \\
\textbf{Factor $2$} & \multicolumn{2}{r}{---} & \multicolumn{2}{r}{$1.885$ ($1.175$, $3.023$)$^{\ast}$} & \multicolumn{2}{r}{$4.715$ ($2.725$, $8.157$)$^{\ast}$} \\
\textbf{Sex} ($0$---Boy; $1$---Girl) & \multicolumn{2}{r}{---} & \multicolumn{2}{r}{$1.359$ ($0.709$, $2.605$)} & \multicolumn{2}{r}{$2.411$ ($1.129$, $5.146$)$^{\ast}$} \\
\textbf{Race} ($0$---White; $1$---Others) & \multicolumn{2}{r}{---} & \multicolumn{2}{r}{$0.814$ ($0.440$, $1.506$)} & \multicolumn{2}{r}{$0.927$ ($0.481$, $1.786$)} \\
\textbf{School Location} & \multicolumn{2}{r}{---} & \multicolumn{2}{r}{$0.776$ ($0.528$, $1.140$)} & \multicolumn{2}{r}{$0.664$ ($0.425$, $1.038$)} \\
\textbf{School Type} ($0$---Public; $1$---Private) & \multicolumn{2}{r}{---} & \multicolumn{2}{r}{$2.149$ ($0.593$, $7.793$)} & \multicolumn{2}{r}{$2.929$ ($0.581$, $14.772$)} \\
\hline
\hline
\end{tabular}
\label{tbl:GMM_reduction}
\begin{tablenotes}
\small
\item[1] Intercept was defined as mathematics IRT scores at 60-month old in this case.
\item[2] $^{\ast}$ indicates statistical significance at $0.05$ level.
\item[3] OR ($95\%$ CI) indicates Odds Ratio ($95\%$ Confidence Interval).
\item[4] We set Class $1$ as the reference group.
\end{tablenotes}
\end{threeparttable}}
\end{table}

\end{document}